\documentclass[11pt,tightenlines,eqsecnum,floats,aps,nofootinbib,prd,shownopacs,floatfix]{revtex4}
\usepackage[english]{babel}
\usepackage[utf8]{inputenc}
\usepackage{amsmath,amsfonts,amssymb,extarrows,nicefrac}
\usepackage[Symbolsmallscale]{upgreek}
\usepackage{xcolor}
\usepackage{graphicx}
\usepackage[hyperfootnotes=false]{hyperref} 
\usepackage{bbm}
\usepackage{pdfpages}
\usepackage{makecell}
\usepackage{mathtools}
\usepackage{mathrsfs}
\usepackage{ stmaryrd } 
\usepackage[scr=dutchcal]{mathalfa}
\usepackage{textgreek}

\newtheorem{theorem}{Theorem}
\newtheorem{lemma}{Lemma}

\newcommand{\kdiff}[0]{k_{\rm d}}

\newcommand{\Cn}[0]{C_{\rm norm}}

\newcommand{\Ptg}{\Psi^t_{\theta_0,p}}

\newcommand{\infsum}[1]{\sum_{#1=-\infty}^{\infty}} 

\newcommand{\norm}[1]{\left\lvert\!\left\lvert #1 \right\rvert\!\right\rvert} 
\usepackage{tikz}
\usetikzlibrary{fit}
\usetikzlibrary{patterns}
\usetikzlibrary{decorations.pathreplacing}
\tikzset{%
  highlight/.style={rectangle,rounded corners, fill=white, pattern color=gray, pattern=crosshatch dots, fill opacity=0.4,draw,thin,inner sep=0pt}
}

\newcommand{\be}{\begin{equation}}
\newcommand{\ee}{\end{equation}}

\let\originalleft\left
\let\originalright\right
\renewcommand{\left}{\mathopen{}\mathclose\bgroup\originalleft}
\renewcommand{\right}{\aftergroup\egroup\originalright}
\newcommand{\lr}[1]{\left(#1\right)} 
\newcommand{\lrabs}[1]{\left|#1\right|} 
\newcommand{\e}[1]{\ensuremath{\mathrm{e}^{#1}}}
\newcommand{\es}[1]{\;\! \ensuremath{\mathrm{e}^{#1}}} 

\newcommand{\abs}[1]{\ensuremath{\left|#1\right|}}
\renewcommand{\d}{\mathrm{d}}

\newcommand{\inv}{{}^{-1}} 

\newcommand{\infintpi}[1]{\int_{-\infty}^{\infty}\frac{\d#1}{\sqrt{2\pi}}\,}

\renewcommand{\i}{\ensuremath{\mathrm{i}}} 
\newcommand{\ih}{\frac{\i}{\hbar}} 
\newcommand{\kchf}[1]{\ensuremath{{}_1\text{F}_1\!\left(#1\right)}} 
\newcommand{\Pqp}{\Psi^{\hbar}_{q,p}} 
\newcommand{\shs}{\frac{\sigma^2}{\hbar^2}} 
\newcommand{\sh}{\frac{\sigma}{\hbar}} 
\newcommand{\qr}{{\hat q}^{i_0}_{I_0}(v,r)} 

\newcommand{\V}{{\hat V}} 
\newcommand{\Q}{{\hat Q}} 

\newcommand{\rhalf}{\frac{r}{2}}

\renewcommand{\pi}{\text{\textpi}} 

\begin{document}

\title{{Coherent States on the Circle: Semiclassical Matrix Elements in the Context of Kummer Functions and the Zak transformation}}

\author{Kristina Giesel}
\thanks{kristina.giesel@gravity.fau.de}
\author{David Winnekens}
\thanks{david.winnekens@fau.de}
\affiliation{Institute for Quantum Gravity, Theoretical Physics III \\ Department of Physics, FAU Erlangen-N\"urnberg, \\ \mbox{Staudtstr.~7}, 91052 Erlangen, Germany}
\begin{abstract}
We extend former results for coherent states on the circle in the literature in two ways. On the one hand, we show that expectation values of fractional powers of momentum operators can be computed exactly analytically by means of Kummer's confluent hypergeometric functions. Earlier, these expectation values have only been obtained by using suitable estimates. On the other hand, we consider the Zak transformation not only to map harmonic oscillator coherent states to coherent states on the circle as it has been discussed before, but we also use the properties of the Zak transformation to derive a relation between matrix elements with respect to coherent states in $L_2(\mathbbm{R})$ and $L_2(S_1)$. This provides an alternative way for computing semiclassical matrix elements for coherent states on the circle. In certain aspects, this method simplifies the semiclassical computations in particular if one is only interested in the classical limit, that is the zeroth order term in the semiclassical expansion. 
\end{abstract}

\maketitle


\section{Introduction}
In this paper, we discuss coherent states for a particle on a circle that have for instance been discussed in earlier work in \cite{Tolar,Kowalski_1996,Olmo,Kastrup:2005xb,Bahr:2006kb} and references therein. In  \cite{Tolar,Kowalski_1996,Olmo,Kastrup:2005xb}, coherent states in the Hilbert space $L_2(S_1)$ were constructed by means of the so-called Zak transformation \cite{Zak1967}, whereas in \cite{Bahr:2006kb} complexifier coherent states \cite{GCS1} for the group U(1) were used, leading finally to the same kind of coherent states. These complexifier coherent states have been introduced in the framework of loop quantum gravity for the group SU(2) and their properties have been analysed in \cite{GCS2,GCS3,GCS4}. Further applications of these coherent states can for instance be found in \cite{Towards1,Towards2,Brunnemann1,Brunnemann2,AQG2,AQG3}.
Expectation values for elementary operators like (integer powers) of the holonomy as well as the momentum operator with respect to coherent states in $L_2(S_1)$, also called semiclassical expectation values, have been computed in \cite{Tolar,Kowalski_1996,Olmo,Kastrup:2005xb,Bahr:2006kb}. 
These semiclassical expectation values can be understood as an expansion in a classicality parameter, denoted by $t$ in our work. For the standard harmonic oscillator coherent states, this classicality parameter can be identified with $\hbar/(m\omega)$. One is then interested in the classical limit of the expectation values, that is when $t$ is sent to zero. In case that a set of coherent states provide an appropriate description of the semiclassical sector of the given quantum theory, we expect that at least for the elementary operators the quantum theory is built from, the classical limit (zeroth order in the semiclassical parameter $t$) agrees with the corresponding classical theory. Such an analysis allows to check whether, for a given operator in the quantum theory, the considered coherent states are suitable. In \cite{Olmo,Kowalski_1996,Kastrup:2005xb}, expectation values with respect to coherent states in $L_2(S_1)$ were expressed in terms of Jacobi's Theta function and its derivatives, which naturally occurs if one applies the Zak transform onto a Gaussian, because the Theta function is the image of a Gaussian under the Zak transformation. 
\\
\\
In our work, we extend these former results into two directions. On the one hand, we generalise the computation of semiclassical expectation values from integers powers of momentum operators to fractional powers. The motivation for his comes from loop quantum gravity and loop quantum cosmology respectively where operators like the square root of the determinant of momentum operators play a pivotal role when the dynamics is quantised. This can be done by using Kummer's confluent hypergeometric functions of the first and second kind. A basic result \cite{Pichler} we will rely on is the fact that the Kummer functions of the first and second kind are mapped into each other under a Fourier transformation if their parameters are adjusted accordingly. This allows to compute those expectation values completely analytical without any estimates for the integrals involved. In a further step, one can use the well-known asymptotic expansions of Kummer's functions in order to obtain an expansion in terms of $\hbar$ or any other classicality parameter. This also allows to compute the classical limit, the lowest order of that expansion, for this kind of expectation values. 
\\
The second direction we will explore is that we consider the Zak transformation not only to obtain coherent states in $L_2(S_1)$ as it has been done in \cite{Olmo,Kowalski_1996,Kastrup:2005xb} but also in the context of computing semiclassical expectation values. By using the basic properties of the Zak transformation, we can show that there exists a very simple relation between the semiclassical matrix elements in $L_2(\mathbbm{R})$ and $L_2(S_1)$. For a given operator on $L_2(S_1)$ (or a suitable domain thereof), the semiclassical matrix elements can be understood as a Fourier series with Fourier coefficients made from the corresponding matrix elements in $L_2(\mathbbm{R})$, that involve the counterpart of the operator on $L_2(\mathbbm{R})$ as well as a translation operator. In particular, this means that any semiclassical matrix element in $L_2(S_1)$ is completely determined by these corresponding matrix elements in $L_2(\mathbbm{R})$. Interestingly, the leading order term, that is the limit in which the semiclassical parameter vanishes, exactly agrees with the semiclassical result obtained in $L_2(\mathbbm{R})$. The latter is just a consequence of the unitarity of the Zak transformation. 
This relation, obtained in section~\ref{sec:RelQMU1}, provides an alternative way for computing semiclassical matrix elements and expectation values respectively. It might also allow to reconsider the techniques from a different angle that have been used in the context of U(1) complexifier coherent states in \cite{GCS2,GCS3,Towards1,Towards2,Bahr:2006kb} in order to estimate semiclassical expectation values and to obtain the classical limit. Although we will restrict to the one-dimensional case in this article, the Zak transformation, and thus also the results presented here, can be easily generalised to higher finite dimensional systems. As a further and more complex application of the techniques developed in this paper, we will apply them to U$(1)^3$ coherent states, which are often used as a toy model for loop quantum gravity, in a companion article \cite{Giesel:2021yop}. There, we will be mainly interested in computing the semiclassical expectation values of dynamical operators as it has for instance been done in \cite{Brunnemann1,Brunnemann2,Towards1,Towards2}. The usage of Kummer functions in this context allows to analytically compute some parts that have been only estimated in earlier work. 
\\
\\
The paper is structured as follows: In section \ref{sec:IntroKummer}, we present the basic properties of Kummer functions that will be used through our work for the benefit of the reader. As a warm up example, we discuss in section \ref{sec:QMCase} fractional powers of momentum operators in the standard quantum mechanical case and compute their semiclassical expectation values analytically by means of Kummer functions and their Fourier transforms respectively. Moreover, as a comparison with the result via Kummer functions, we apply in section \ref{sec:QMAQG} a different technique of computing fractional powers of operators in a more general context, called the AQG-III algorithm \cite{AQG3}. Next, we move on to quantum mechanics on a circle in section \ref{sec:ZakTrafo}. Since, as in \cite{Bahr:2006kb}, we come from the complexifier coherent states, after a brief introduction to the Zak transformation, we apply it to the heat kernel, which is more convenient in this context, and not to the harmonic oscillator coherent states directly. For this purpose, we use former work from \cite{Neretin} and as expected we end up with the same coherent states for $L_2(S_1)$. Given this set of coherent states, we compute semiclassical expectation values of operators involving fractional powers of the momentum operators in the remaining part of section \ref{sec:ZakTrafo}.
The relation between semiclassical matrix elements in $L_2(\mathbbm{R})$ and $L_2(S_1)$ is derived and discussed in section \ref{sec:RelQMU1}. As an application of this relation, we recompute a couple of semiclassical matrix elements and expectation values respectively and show that we obtain the correct results. Before we finally conclude in section \ref{sec:Concl}, we discuss in section \ref{sec:SolHeat} based on results obtained in \cite{SelfSimHeat} that the heat equation can be transformed into Kummer's differential equation for a specific choice of the parameters in the Kummer functions and thus the Kummer functions involved in our work can be understood as solutions of the heat equation for certain choices of boundary data. 

\section{Brief introduction to Kummer's confluent hypergeometric functions}
\label{sec:IntroKummer}
Confluent hypergeometric functions --- also called Kummer functions --- are solutions to Kummer's differential equation \cite{Kummer}
\be
z\frac{\d^2w}{\d z^2}+(b-z)\frac{\d w}{\d z}-aw=0 .
\ee
The two independent solutions are called Kummer functions of the first and second kind respectively and we will denote them by $\kchf{a,b,z}$ and $U(a,b,z)$ respectively\footnote{$\kchf{a,b,z}$ is denoted by $ \varphi(\alpha,\beta,x) $ in eq. 1. in \cite{Kummer} and often it is also denoted by $M(a,b,z)$. $U(a,b,z)$ is also called Tricomi's function, after Francesco Tricomi who introduced them in~\cite{Tricomi}.}:
\begin{align}
    \kchf{a,b,z} &\coloneqq \sum_{n=0}^\infty \frac{(a)_n}{(b)_n n!} z^n \ \text{and} \label{eq:defKCHF}\\
    U(a,b,z) &\coloneqq \frac{\Gamma(1-b)}{\Gamma(1+a-b)}\kchf{a,b,z} + \frac{\Gamma(b-1)}{\Gamma(a)}z^{1-b} \kchf{1+a-b,2-b,z}. \label{eq:defU}
\end{align}
Therein, $(a)_n$ denotes the Pochhammer symbol or raising factorial\footnote{Note that the raising factorial is sometimes also denoted by $a^{(n)}$. To make things even worse, Pochhammer himself used $(a)_n$ for the binomial coefficient $\binom{a}{n}$ and $[a]_n^+$ for the raising factorial~\cite[p.\,80-81]{Pochhammer}. We use the above notation as it became the established standard for hypergeometric functions.}
\begin{align}
    (a)_0 & = 1 , \nonumber \\
    (a)_1 & = a \ \text{ and}\nonumber \\
    (a)_n & = a (a+1) (a+2)\cdots(a+n-1) .
\end{align}
A vast amount of elementary functions can be rewritten in terms of Kummer functions for an appropriate choice of the parameters $a,b,c$, such as  $\kchf{0,b,z}=1$, $\kchf{a,a,z}=\e{z}$, $U(-\frac{n}{2},\frac{1}{2},z^2)=2^nH_{n}(z)$  with the Hermite polynomials $H_n(z)$ as well as further relations to Bateman's function, Bessel functions, Laguerre polynomials etc. as can for instance be found in ~\cite{abramowitz+stegun,dlmf}. The Kummer (confluent hypergeometric) functions of the first kind, $\kchf{a,b,c}$, can also be understood as a special limit of the ordinary (or Gaussian) hypergeometric functions ${}_2\text{F}_1\left(a,c;b;z\right) \coloneqq \sum_{n=0}^\infty \frac{(a)_n(b)_n}{(c)_n n!} z^n$ via
\be
\kchf{a,b,z}=\lim_{c\to\infty}{}_2\text{F}_1\!\left(a,c;b;\frac{z}{c}\right) .
\ee
$\kchf{a,b,z}$ is an entire function in $a$ and $z$ and a meromorphic one in $b$ since it has poles at $b=-n$ with $n\in\mathbbm{N}_0$, whereas $U(a,b,z)$ is entire in $a$ and $b$ except in its branch point at $z=0$. Note that the function $\kchf{a,b,z}/\Gamma(b)$ --- sometimes denoted by an upright M$(a,b,z)$ --- is entire in $a,b$ and $z$.
A special property of both solutions is that they satisfy certain Kummer transformations given by
\begin{align}
\label{eq:KummerTrafo}
    \kchf{a,b,z} &= \e{z}\,\kchf{b-a,b,-z} \ \text{ and} \nonumber \\
    U(a,b,z)     &=  z^{1-b}U(a-b-1,2-b,z)
\end{align}
that we will use later in this work.

To conclude this chapter, we note that many contradictory variations of how to name and hence differ between $\kchf{a,b,z}$, $U(a,b,z)$, M$(a,b,z)$ and ${}_2\text{F}_1\left(a,c;b;z\right)$ exist in the literature. We will mainly use the phrase \textit{Kummer('s) functions} or the abbreviation KCHF for \textit{Kummer's confluent hypergeometric functions} for both $\kchf{a,b,z}$ and $U(a,b,z)$ since we will only be using these two and the context should always clarify which one we refer to at the time.

\subsection{The Fourier transformation of Kummer functions}
The analytical computations of the expectations values with respect to coherent states are heavily based on a theorem from \cite{Pichler}, where it is proven that the Kummer functions of the first and second kind are in certain sense dual to each other under Fourier transformation. We use the following definition of the Fourier transformation and its inverse in one dimension:
\begin{equation*}
 {\cal F}[f](k)\coloneqq \hat{f}(k)=\frac{1}{\sqrt{2\pi}}\int\limits_{\mathbbm{R}} \d x\, f(x)\es{-\i kx} \  \text{ and } \  
 {\cal F}^{-1}[\hat{g}](x)\coloneqq g(x)=\frac{1}{\sqrt{2\pi}}\int\limits_{\mathbbm{R}}\, \d k \, \hat{g}(k)\es{\i kx},
\end{equation*}
where ${\cal F}[f]$ and $\hat{f}$ denote the Fourier transform of $f$ and we will use both notations in the remaining part of the article. 
The aforementioned theorem from \cite{Pichler} then reads:
\begin{theorem}
\label{ThPichler}
(Fourier transform of Kummer’s functions). Kummer’s functions are symmetric with respect to Fourier transformation. Let $x,k\in\mathbbm{R}$, we have for ${\rm Re}(b-a)>0$
\begin{align}
   {\cal F}\left(\e{-x^2}U\left(a,b,x^2\right)\right) &=\frac{1}{\sqrt{2}}\frac{\Gamma\left(\frac{3}{2}-b\right)}{\Gamma\left(a-b+\frac{3}{2}\right)}\es{-\frac{k^2}{4}} {}_1\mathrm{F}_1\left(a,a+\frac{3}{2}-b,\frac{k^2}{4}\right) \ \mathrm{ and} \label{eq:FourierU}\\
   {\cal F}\left(\e{-x^2}{}_1\mathrm{F}_1\left(a,b,x^2\right)\right) &=\frac{1}{\sqrt{2}}\frac{\Gamma(b)}{\Gamma(b-a)}\es{-\frac{k^2}{4}}U\lr{a,a+\frac{3}{2}-b,\frac{k^2}{4}}. \label{eq:Fourier1F1}
\end{align}
\end{theorem}
The proof of this theorem in \cite{Pichler} is based on an expansion of Kummer's functions in terms of associated Laguerre polynomials, which necessitates the restriction ${\rm Re}(b-a)>0$ (cf. theorem 1 of~\cite{Pichler}). We need a slight modification of theorem \ref{ThPichler} for some of the integrals involved in our work. For this purpose, we use the modulation property of the Fourier transform, that is 
${\cal F}(\e{\i k_0 x}f(x))=\hat{f}(k-k_0)$. Carried over to Kummer's functions we obtain
\begin{lemma}
\label{LemModFourier}
 Let $x,k\in\mathbbm{R}, {\rm Re}(b-a)>0$ and $\alpha\in\mathbbm{C}$. Then, we have
\begin{align}
  {\cal F}\left(\e{-x^2}U\left(a,b,x^2\right)\e{\i \alpha x}\right)&= \frac{1}{\sqrt{2}}\frac{\Gamma\left(\frac{3}{2}-b\right)}{\Gamma\left(a-b+\frac{3}{2}\right)}\es{-\frac{(k-\alpha)^2}{4}} {}_1\mathrm{F}_1\left(a,a+\frac{3}{2}-b,\frac{(k-\alpha)^2}{4} \right) \ \mathrm{and}\\  
   {\cal F}\left(\e{-x^2} {}_1\mathrm{F}_1\left(a,b,x^2\right)\e{\i \alpha x}\right) &=
   \frac{1}{\sqrt{2}}\frac{\Gamma(b)}{\Gamma(b-a)}\es{-\frac{(k-\alpha)^2}{4}}U\lr{a,a+\frac{3}{2}-b,\frac{(k-\alpha)^2}{4}}.
\end{align}
\end{lemma}
Given theorem \ref{ThPichler} and the fact that the Kummer functions can be analytically continued to the complex plane except in the branch point $z=0$ of $U(a,b,z)$, this lemma can be as easily proved as the modulation property of the Fourier transform itself, thus we will not present it here. These results on the Fourier transformation of Kummer's functions will play a pivotal role in our analytic computations. 
\subsection{The asymptotic expansion for large arguments of Kummer functions}

One of the properties of Kummer functions that we will often make use of is the asymptotic expansion for large arguments $|z|\to\infty$~\cite{abramowitz+stegun}:
\begin{align}
\kchf{a,b,z} \overset{\lrabs{z}\to \infty}{\approx} \Gamma(b) & \left[ 
\frac{\e{\pm\pi\i a}z^{-a}}{\Gamma(b-a)} \sum_{n=0}^{\infty}
\frac{(a)_n(1+a-b)_n}{n!}(-z)^{-n} +\right. \nonumber \\
& \left.\quad\! +\frac{\e{z} z^{a-b}}{\Gamma(a)} \sum_{n=0}^{\infty} \frac{(b-a)_n(1-a)_n}{n!}z^{-n} \right] . \label{expansion}
\end{align}
Therein, the minus sign in $\exp\lr{{\pm\pi\i a}}$ is chosen if $z$ lies in the right half plane~\cite{abramowitz+stegun}.
For the Kummer function of the second kind, the asymptotic expansion for large arguments reads
\begin{equation}
    U(a,b,z) \overset{\lrabs{z}\to \infty}{\approx} z^{-a} \sum_{n=0}^\infty \frac{(a)_n (1+a-b)_n}{n!}(-z)^{-n} .
\end{equation}{}
Note that we will from now on always use the symbol $\approx$ to mark the usage of this asymptotic expansion, or any other series expansion that is only true if some parameter approaches a certain, declared value.

\section{Warmup Example: Kummer functions and coherent states in quantum mechanics}
\label{sec:QMCase}
As a warm-up exercise, we consider the expectation value of fractional powers of the momentum operator with respect to coherent states used in one-dimensional quantum mechanics. Afterwards, we will apply this result to a toy model operator of the form $\e{\i\alpha\hat x}\left[\e{-\i\alpha\hat{x}},\sqrt[4]{\lrabs{\hat p}^3}\right]$, which mimics certain dynamical operators within loop quantum gravity. In particular, we want to compare the results we obtain using Kummer's functions with the one we get when employing semiclassical perturbation theory in the form introduced in \cite{AQG3}.

We consider the position representation and normalised coherent states of the form
\be
\label{eq:CSQM}
\Psi_{\text{coh}} = \Pqp = \frac{1}{\sqrt[4]{\pi}\sqrt{\sigma}}
\es{-\frac{(x-q)^2}{2\sigma^2}}\es{\ih px} \ ,
\ee
where the subscript $q,p$ denotes the state being peaked around $(q,p)$ in phase space and $\sigma = \sqrt{\frac{\hbar}{m\omega}}$ is the width of the coherent state in phase space. The expectation value we want to compute is given by
\begin{equation}
\langle |\hat{p}|^r\rangle_{\Pqp} = 
\langle\Pqp \, |\,|\hat{p}|^r\,|\, \Pqp\rangle 
= \int\limits_{-\infty}^\infty \d k \, {\cal F}[\overline{\Pqp}](k) \, |k|^{r} \, {\cal F}[\Pqp](k),\nonumber
\end{equation}
with $r$ being a rational number. In order that the integral is well defined we need $r>-1$. To ensure that also the norm of the state $|\hat{p}|^r\Pqp$ stays finite we further restrict $r>-\frac{1}{2}$. Above, we inserted a resolution of identity in terms of generalised momentum eigenstates and used that the momentum operator acts diagonally on them. Now, using that a Gaussian is self-reciprocal with respect to the Fourier transform and the modulation property of the Fourier transform, we end up with
\begin{equation*}
\langle |\hat{p}|^r\rangle_{\Pqp}=\frac{\sigma}{\hbar\sqrt{\pi}}\int\limits_{-\infty}^\infty \d k \left(k^2\right)^{\frac{r}{2}}\e{-\frac{\sigma^2}{\hbar^2}(k-p)^2}
=\frac{\sigma}{\sqrt{\pi}\hbar}\e{\left(\frac{\sigma}{\hbar}p\right)^2}\int\limits_{-\infty}^\infty \d k \left(k^2\right)^{\frac{r}{2}}
\e{-\left(\frac{\sigma}{\hbar}k\right)^2} \e{\i \left(-2\i \frac{\sigma}{\hbar} p\right)\frac{\sigma}{\hbar} k}.
\end{equation*}
In order to express the above integral in terms of Kummer's functions, we use the following property of $U(a,b,z)$:
\begin{eqnarray*}
|k|^r&=&(k^2)^{\frac{r}{2}}=\left(\frac{{\hbar}}{\sigma}\right)^{\!r}\left(
\left(\frac{\sigma}{{\hbar}}k\right)^2\right)^{\frac{r}{2}}  = \left(\frac{{\hbar}}{\sigma}\right)^{\!r} U\lr{-\frac{r}{2},-\frac{r}{2}+1,\left(\frac{\sigma}{{\hbar}}k\right)^2}.   
\end{eqnarray*}
Hence, we can rewrite the integral above as
\begin{equation}
\label{eq:ExpKummer}
\langle |\hat{p}|^r\rangle_{\Pqp}=\frac{\sigma}{\hbar{\sqrt{\pi}}}\e{\left(\frac{\sigma}{\hbar}p\right)^2}
\left(\frac{{\hbar}}{\sigma}\right)^r
\int\limits_{-\infty}^\infty \d k\,  U\lr{-\frac{r}{2},-\frac{r}{2}+1,\left(\frac{\sigma}{{\hbar}}k\right)^2}
\e{-\left(\frac{\sigma}{\hbar}k\right)^2}\e{\i \left(-2\i \frac{\sigma}{{\hbar}}p\right)\frac{\sigma}{\hbar}k}.
\end{equation}
Now, considering the scaling property of the Fourier transform as well as theorem~\ref{ThPichler}, we can express the Kummer function $U(\ldots)$ involved in the integral together with the Gaussian as a Fourier transform of $\kchf{\ldots}$ --- cf.~\eqref{eq:Fourier1F1}. Explicitly, we obtain
\begin{equation*}
 \frac{\sqrt{2}\Gamma\left(\frac{r+1}{2}\right)}{\Gamma\left(\frac{1}{2}\right)}\frac{\hbar}{2\sigma}{\cal F}\left(\e{-\left(\frac{\hbar x}{2\sigma}\right)^2}\kchf{-\frac{r}{2},\frac{1}{2},\left(\frac{\hbar x}{2\sigma}\right)^2}\right)=
 U\lr{-\frac{r}{2},-\frac{r}{2}+1,\left(\frac{\sigma}{{\hbar}}k\right)^2}
\e{-\left(\frac{\sigma}{{\hbar}}k\right)^2}.
\end{equation*}
Using this as well as the definition of the inverse Fourier transformation, the expectation value for $r>-\frac{1}{2}$ finally reads
\begin{align}
\label{eq:KCHFpr}
\langle\Pqp \, |\,|\hat{p}|^r\,|\, \Pqp\rangle & = 
\frac{\Gamma\left(\frac{r+1}{2}\right)}{\Gamma\left(\frac{1}{2}\right)}
 \left( \frac{{\hbar}}{\sigma}\right)^{\!r} \es{\frac{\sigma^2}{\hbar^2}p^2}
 {\cal F}^{-1}\left[{\cal F}\left[\e{-\left(\frac{\hbar x}{2\sigma}\right)^2}\kchf{-\frac{r}{2},\frac{1}{2},\left(\frac{\hbar x}{2\sigma}\right)^2}\right]\right]
  \nonumber \\
 & =  \frac{\Gamma\left(\frac{r+1}{2}\right)}{\Gamma\left(\frac{1}{2}\right)}
\left( \frac{{\hbar}}{\sigma}\right)^{\!r} \es{\frac{\sigma^2}{\hbar^2}p^2}
 \e{-\left(\frac{\sigma  p}{\hbar}\right)^2}\kchf{-\frac{r}{2},\frac{1}{2},-\left(\frac{\sigma p}{\hbar}\right)^2}  \nonumber \\
& = 
 \frac{\Gamma\left(\frac{r+1}{2}\right)}{\sqrt{\pi}}
\left( \frac{{\hbar}}{\sigma}\right)^{\!r}\kchf{-\frac{r}{2},\frac{1}{2},-\left(\frac{\sigma p}{\hbar}\right)^2}.  
\end{align}
In the first line, we absorbed the factor of $1/\sqrt{2\pi}$ into the inverse Fourier transformation and, in the last line, we used that $\Gamma\left(\frac{1}{2}\right)=\sqrt{\pi}$. Thus, we managed to compute the expectation value analytically without using any estimations or Taylor expansion on the way as it has been done for these kind of operators in the existing literature \cite{Brunnemann1,Brunnemann2,Towards1,Towards2,AQG3}. As a crosscheck, we realise that for the choice of $r=0$ the result equals $1$ because $\kchf{0,\frac{1}{2},-\left(\frac{\sigma p}{\hbar}\right)^2}=1$ and the normalisation of the coherent states is recovered. To apply lemma~\ref{LemModFourier} about the Fourier transform of Kummer functions, we need to require ${\rm Re}(b-a)>0$ and note that this is consistent with our restriction of $r>-\frac{1}{2}$ because for $\kchf{-\frac{r}{2},\frac{1}{2},.}$ this carries over to ${\rm Re}\left(\frac{1}{2}+\frac{r}{2}\right)>0$ being fulfilled.

We can now also proceed from the result above by applying the asymptotic expansion for large arguments of the KCHF according to (\ref{expansion}). Note that in this scenario of the quantum mechanical harmonic oscillator coherent states, we have $\hbar$ as a natural quantity that supplies tininess. However, this procedure is of course applicable to the presence of other classicality parameters too. With the KCHF's argument $z=-\left(\frac{\sigma p}{\hbar}\right)^2 = -\frac{p^2}{\hbar m\omega}$, we obtain
\begin{align}
\langle\Pqp \, |\,|\hat{p}|^r\,|\, \Pqp\rangle & \approx \frac{\Gamma\lr{\frac{r+1}{2}}}{\sqrt{\pi}}\lr{\frac{\hbar}{\sigma}}^{\!r}   \Gamma\lr{\tfrac{1}{2}} \left[ 
\frac{\e{\mp\pi\i \frac{r}{2}} {\lr{-\frac{p^2}{\hbar m\omega}}}^{\frac{r}{2}}}{\Gamma\lr{\frac{r+1}{2}}} \sum_{n=0}^{\infty}
\frac{(-\frac{r}{2})_n\lr{\frac{1-r}{2}}_n}{n!}\lr{\frac{\hbar m\omega}{p^2}}^{n} +\right. \nonumber \\
& \left.\quad +\frac{\e{-\frac{p^2}{\hbar m\omega}} \lr{-\frac{p^2}{\hbar m\omega}}^{-\frac{1+r}{2}}}{\Gamma(-\frac{r}{2})} \sum_{n=0}^{\infty} \frac{\lr{\frac{1+r}{2}}_n(1+\frac{r}{2})_n}{n!}\lr{-\frac{\hbar m\omega}{p^2}}^{n} \right] \nonumber\\
& \approx \lrabs{p}^r \sum_{n=0}^{\infty} \frac{\lr{-\frac{r}{2}}_n \lr{\frac{1-r}{2}}_n}{n!}\lr{\frac{\hbar m\omega}{p^2}}^n\nonumber\\
& \approx \abs{p}^r \lr{1-\frac{r(1-r)}{4}\frac{\hbar m \omega}{p^2} + \mathcal{O}\lr{\hbar^2}}, \label{eq:QMprExpansion}
\end{align}
where we first of all neglected the second sum in the expansion's square bracket due to the Gaussian prefactor damping it to $\mathcal{O}\lr{\hbar^{\infty}}$.\footnote{This is a feature always happening when performing the asymptotic expansion: one of the sums vanishes. Which one it is depends on whether we have another inverse Gaussian in front of the expansion or not.} With the KCHF's argument being negative and $a = -\rhalf$, we have to choose the minus sign in the first sum's exponential function prefactor, yielding $\e{-\pi\i\frac{r}{2}}\lr{-1}^{\rhalf} = \lr{-1}^{-\rhalf}\lr{-1}^{\rhalf} = 1$ and no imaginary part remains. Note that the expression above is in fact not an estimate but indeed a series expansion calculable up to arbitrary order for semiclassical expectation values.

\subsection{The AQG-III-algorithm}\label{sec:QMAQG}

We just illustrated how to compute semiclassical expectation values of fractional powers of the momentum operator in standard quantum mechanics by means of KCHFs. Ultimately, we're interested in performing similar calculations on a more complex operator within loop quantum gravity (LQG), the class of so-called ${\hat q}^{i_0}_{I_0}(r)$-operators\footnote{Note that we use the U$(1)^3$-description of loop quantum gravity. While LQG is actually a SU(2)-theory,~\cite{GCS2,GCS3} showed that the replacement of SU(2) by U$(1)^3$ does not change the outcome of semiclassical calculations qualitatively.}. They play a crucial role when studying the dynamics of loop quantum gravity and many effort is put into estimating expectation values of them~\cite{Brunnemann1,Brunnemann2,Towards1,Towards2}. Written down explicitly, they are of the form
\begin{equation}
    \qr = {\hat h}^{i_0}_{I_0}\left[\left({\hat h}{}^{i_0}_{I_0}\right)^{-1},{\hat V}_v^r\right] , \label{eq:qr}
\end{equation}
where $ {\hat h}^{i_0}_{I_0}$ is the holonomy operator acting on edge $e_{I_0}$ and U(1)-copy $i_0$ by increasing the state's U(1)-charge $n^{i_0}_{I_0}$ by 1 and ${\hat V}_v^r$ is the volume operator to the power of $r\in\mathbbm{Q}$:
\begin{equation}
    {\hat V} = \kappa_{\mathrm{V}}  \sum_{v} {\hat V}_{v} = \kappa_{\mathrm{V}} \sum_{v} \sqrt{\left|{\hat Q_{v}}\right|} .
\end{equation}
In U$(1)^3$-LQG, the coherent states are essentially products of Gaussians in the charges $n^i_I$ for all edges $e_I$ of the graph the state is defined on and the three U(1)-copies $i=1,2,3$ per such edge. The points of the graph where the edges meet and the volume operator ${\hat V}_v$ is evaluated at are called vertices $v$.

The computational problem now hides in the operator ${\hat Q_{v}}$, which collects momentum-like contributions from all triples of edges that meet at the vertices $v_I$ within the region one wants to determine the volume of. As these contributions are determinants of three edges and their relative U(1)-copies, taking the root of the absolute value of the sum of those yields a hard-to-handle, non-polynomial  operator. This is the reason why one tries to find approximations or estimates in order to obtain for example bounds of its expectation values w.r.t. coherent states.

One way of determining semiclassical expectation values of this kind of operators is offered by algebraic quantum gravity,~\cite{AQG1,AQG2,AQG3,AQG4}. In~\cite{AQG3}, it is shown that one can replace the volume operator's root within the expectation value w.r.t. $\Psi$ in such a way that one obtains a power series in the classicality parameter $t$ and expectation values of only integer powers of ${\hat Q_{v}}$,
\begin{align}
    \V^{4q}_{v} \mapsto \lr{\langle \Q_{v} \rangle_{\Psi}}^{2q}\lr{1+\sum_{n=1}^{2k+1}(-1)^{n+1}\,\frac{q(1-q)\cdots(n-1-q)}{n!}\lr{\frac{\Q_{v}^2}{\langle\Q_{v}\rangle_{\!\Psi}^2}-1}^{\!\!\!n}\,} , \label{eq:AQG3}
\end{align}
while the error one makes is of order $\hbar^{k+1}$.

We will now apply this procedure to the maybe most drastic simplification, where we carry the whole mechanism over to standard quantum mechanics and adapt the ${\hat q}^{i_0}_{I_0}(v,r)$-operators' action therein as
\begin{align}
    {\hat q}_{\text{qm}}(r) = \e{\i\alpha\hat x}\left[\e{-\i\alpha\hat x},\lrabs{\hat p}^r\right] = \lrabs{\hat p}^r - \e{\i\alpha\hat x} \lrabs{\hat p}^r \e{-\i\alpha\hat x} .
\end{align}
This means they evaluate the $r$-th root of the momentum operator's absolute value not only on the (coherent) state itself, but also on ${\tilde\Psi}^{\hbar}_{q,p} = \e{-\i\alpha\hat x}\Pqp$ and then return the difference of the both. As we will see later, the action of $\e{-\i\alpha\hat x}$ acts like an infinitesimal shift and taking the difference of the two expectation values with additionally dividing by $\hbar$, we eventually get the derivative of $\langle\lrabs{\hat p}^r\rangle_{\Pqp}$ in the semiclassical limit $\hbar\to 0$.

As a special case, we will focus on $r=\frac{1}{2}$ in \eqref{eq:qr} as the square root of the volume operator is a frequently appearing object and also the most basic choice for a root. As mentioned already, the ${\hat Q}_v$-operator is an operator that collects momentum-like contributions from triples of edges. Hence, we mimic it as ${\hat Q}_v \mapsto {\hat p}^3$ in the quantum mechanics framework.

If we are then interested in including up to $\hbar^2$-corrections, we need to evaluate \eqref{eq:AQG3} up to $k=1$, which involves expectation values of ${\hat p}$ up to ${\hat p}^{18}$. As these are all integer powers of the momentum operator, expectation values thereof can be computed by the standard methods. In the end, we obtain
\begin{align}
\langle \e{\i\alpha\hat x}\left[\e{-\i\alpha\hat x},\sqrt[4]{\lrabs{\hat p}^3}\right]\rangle_{\Pqp} & \approx \frac{3}{4} \frac{\alpha\lrabs{p}^{\frac{3}{4}}}{p}\hbar + \lr{ \frac{3}{32} \frac{\alpha^2\lrabs{p}^{\frac{3}{4}}}{p^2} + \frac{15}{256} \frac{\alpha m \omega\lrabs{p}^{\frac{3}{4}}}{p^3}}\hbar^2 + \mathcal{O}\lr{\hbar^3}. \label{eq:qmAQG3result}
\end{align}
We can now compare this result's lowest order term with the Poisson bracket of the operators' classical counterparts and see immediately that the two coincide after dividing the first term of~\eqref{eq:qmAQG3result} additionally by $\hbar$.

\subsection{Direct calculations and KCHF}\label{sec:QMdirect}

With the final result of (\ref{eq:KCHFpr}), we can, however, compute this expectation value also directly analytically --- having the previous result via the AQG-III-algorithm as a crosscheck in mind. Choosing accordingly $r=\frac{3}{4}$ in \eqref{eq:KCHFpr}, we immediately obtain
\begin{align}
\langle \sqrt[4]{\lrabs{\hat p}^3} \rangle_{\Pqp} &= \langle\Pqp|\sqrt[4]{\lrabs{\hat p}^3}|\Pqp\rangle \nonumber = \sh\frac{1}{\sqrt{\pi}} \int_{-\infty}^\infty \d k \abs{k}^{\frac{3}{4}} \e{-\shs(k-p)^2}\nonumber \\
& = \frac{1}{\sqrt{\pi}}\left(\frac{\hbar}{\sigma}\right)^{\!\frac{3}{4}} \Gamma\left(\tfrac{7}{8}\right)\kchf{-\frac{3}{8},\frac{1}{2},-\shs p^2} .\label{eq:QMunshiftedp}
\end{align}
Likewise, we get 
\be
\langle \sqrt[4]{\lrabs{\hat p}^3} \rangle_{\e{-\i\alpha \hat x}\Pqp} = \frac{1}{\sqrt{\pi}}\left(\frac{\hbar}{\sigma}\right)^{\!\frac{3}{4}} \Gamma\left(\tfrac{7}{8}\right)\kchf{-\frac{3}{8},\frac{1}{2},-\shs (p-\hbar\alpha)^2} \label{eq:QMshiftedp}
\ee
for the shifted expectation value. Note that we can directly see how $\e{-\i\alpha \hat x}$ acts on the quantum mechanical coherent state by combining it with the last e-function of the state's definition~\eqref{eq:CSQM}. This allows us to understand it as a shift in the momentum: $p \mapsto p-\hbar\alpha$.

Since the arguments of both KCHF go with $-\shs p^2 = -\frac{p^2}{m\omega \hbar}$, i.e. with $\frac{1}{\hbar}$ in at least one of their terms, we can make use of the asymptotic expansion for large arguments, (\ref{expansion}). By that, we get
\begin{align}
\langle\Pqp|& \e{\i\alpha\hat x}\left[\e{-\i\alpha\hat
x},\sqrt[4]{\lrabs{\hat p}^3}\right] |\Pqp\rangle \approx \frac{3}{4}\frac{\alpha\lrabs{p}^{\frac{3}{4}}}{p}\hbar + \lr{\frac{3}{32}\frac{\alpha^2\lrabs{p}^{\frac{3}{4}}}{p^2} + \frac{15}{256}\frac{\alpha m\omega\lrabs{p}^{\frac{3}{4}}}{p^3}}\hbar^2 \nonumber\\
& + \lr{ \frac{5}{128} \frac{\alpha^3\lrabs{p}^{\frac{3}{4}}}{p^3} + \frac{135}{2048} \frac{\alpha^2 m \omega\lrabs{p}^{\frac{3}{4}}}{p^4} + \frac{1755}{32768} \frac{\alpha m^2 \omega^2\lrabs{p}^{\frac{3}{4}}}{p^5}}\hbar^3 +\mathcal{O}\lr{\hbar^4} \ , \label{eq:QMdirectKCHF}
\end{align}
where we performed the expansion analogously to~\eqref{eq:QMprExpansion}, i.e. one sum per expansion is neglected straight away due to the damping Gaussian prefactor. So we see that the direct approach via KCHFs reflects nicely the results of the AQG-III-algorithm while calculating higher correction terms is less laborious.

However, if we want to consider the case $p=0$, we have to proceed differently. First, we insert $p=0$ into~\eqref{eq:QMunshiftedp} \&~\eqref{eq:QMshiftedp}. The KCHF of~\eqref{eq:QMunshiftedp} therefore becomes 1 as its argument vanishes --- $\kchf{a,b,0}=1$. By inserting the very definition of the KCHF,~\eqref{eq:defKCHF}, into~\eqref{eq:QMshiftedp} with $p=0$, we automatically get a power series in $\hbar$:
\begin{align}
    \langle\Pqp| \e{\i\alpha\hat x}\left[\e{-\i\alpha\hat
x},\sqrt[4]{\lrabs{\hat p}^3}\right] |\Pqp\rangle &\stackrel{p=0}{=} \frac{1}{\sqrt{\pi}}\lr{\hbar m\omega}^{\frac{3}{8}} \Gamma\lr{\tfrac{7}{8}} \lr{1 - \sum_{n=0}^{\infty} \frac{\lr{-\frac{3}{8}}_n}{\lr{\frac{1}{2}}_n n!} \lr{-\frac{\alpha\hbar}{m\omega}}^n} \nonumber\\
& = -\frac{3\Gamma\lr{\tfrac{7}{8}}}{4\sqrt{\pi}} \frac{\alpha^2}{\lr{m\omega}^{\frac{5}{8}}}\hbar^{\frac{11}{8}} + \mathcal{O}\lr{\hbar^{\frac{19}{8}}}.
\end{align}
\section{Coherent States on the circle via the Zak transform}
\label{sec:ZakTrafo}
In this section, we want to discuss quantum mechanics on a circle, for which the associated coherent states are U(1) coherent states, as for instance discussed in \cite{Tolar,Kastrup:2005xb,Kowalski_1996,Olmo,Bahr:2006kb}. While \cite{Tolar,Kastrup:2005xb,Kowalski_1996,Olmo} consider the Zak transform to obtain the coherent states on a circle, \cite{Bahr:2006kb} considers them in the context of complexifier coherent states. In this work we want to discuss U(1) coherent states from a slightly different angle and combine the complexifier approach with the Zak transform. For the computation of expectation values with respect to U(1) coherent states we will again use Kummer's function.


The Zak transform, also called Brezin-Weil-Zak transform~\cite{Brezin,Weil,Zak1967}, is a unitary map from $L_2(\mathbbm{R})$ to $L_2(\mathbbm{R}^2/\mathbbm{Z}^2)$, mapping functions from $\mathbbm{R}$ to functions on a torus. It is defined as
\be
{\cal Z}_a \colon L_2(\mathbbm{R})\to L_2(\mathbbm{R}^2/\mathbbm{Z}^2),\quad
f\mapsto {\cal Z}_a[f](x,\zeta)\coloneqq \sqrt{a}\sum\limits_{n=-\infty}^\infty f(x+2\pi na)\e{-2\pi \i  n a\zeta},
\ee
with $x\in [0,2\pi a],\zeta\in [0,1/a]$, were we in an abuse of notation denoted the old and new coordinate by $x$.
The inverse map ${\cal Z}^{-1}_a \colon L_2(\mathbbm{R}^2/\mathbbm{Z}^2)\to L_2(\mathbbm{R})$ is then given by
\be
{\cal Z}^{-1}_a[g](x)\coloneqq\sqrt{a}\int\limits_{0}^{\tfrac{1}{a}} \d\zeta\, g(x,\zeta),
\ee
where $a\not=0\in\mathbbm{R}$. We restrict our discussion to one dimension here, but the Zak transformation can trivially be generalised to $L_2(\mathbbm{R}^n)$. The image of the Zak transform ${\cal Z}_a[f]$ is a function quasiperiodic in $x$ and periodic in $\zeta$, that is for $g={\cal Z}_a[f]$, $ \ell,m\in\mathbbm{Z}$, we have
\begin{equation}
\label{eq:QuasiPer}
g\left(x,\zeta+\tfrac{\ell}{a}\right)=g(x,\zeta),\quad{\rm and}\quad g(x+2\pi a m,\zeta)=\e{2\pi \i ma\zeta }g(x,\zeta).
\end{equation}
Due to this (quasi-)periodicity properties, ${\cal Z}[f]$ is completely determined by its values on the two dimensional rectangle $[0,2\pi a]\times[0,\tfrac{1}{a}]$.

As an example, let us consider the Zak transform of a Gaussian for $a=1$, with ${\cal Z}_{1}\coloneqq{\cal Z}$. We obtain for $f\colon\mathbbm{R}\to\mathbbm{R}, x\mapsto f(x)\coloneqq \e{-\frac{x^2}{4b}}$
\begin{align}
{\cal Z}[f](x,\zeta) &= \sum\limits_{n=-\infty}^\infty \e{-\frac{1}{4b}(x+2\pi n)^2}\e{-2\i \pi n\zeta} \nonumber
\\
&= \e{-\frac{x^2}{4b}}\sum\limits_{n=-\infty}^\infty \e{2\i  n\left(-\pi\zeta +\i 2\pi\frac{x}{4b}\right)}\e{-\frac{(2\pi)^2n^2}{4b}}  \nonumber \\
&= \e{-\frac{x^2}{4b}}\Theta\left(-\pi\zeta+\frac{
2\i\pi x}{4b},\frac{\i\pi}{b}\right),
\end{align}
where we used the definition of Jacobi's third Theta function $\Theta(z,\e{\i\pi\tau})\eqqcolon \Theta(z,\tau)\coloneqq\sum_{n=-\infty}^\infty \e{2\i n z}\e{\i\pi\tau n^2}$ with ${\rm Im}(\tau)>0$.

Below, we will also need the Zak transform combined with a dilatation of the argument of the function, that is for $\gamma>0$, $D_\gamma f(x)=\sqrt{\gamma}f(\gamma x)$. As shown in \cite{Janssen1988}, we have
\be
{\cal Z}[D_\gamma f](x,\zeta)={\cal Z}_{\gamma}[f](\gamma x,\nicefrac{\zeta}{\gamma}), \quad {\cal Z}\coloneqq {\cal Z}_1.
\ee
If we choose in particular $\gamma=a$, we can rewrite the ordinary Zak transform above as
\be
\label{eq:ZakDilat}
{\cal Z}_a[f](x,\zeta)=\sqrt{a}\infsum{n} f(ax+2\pi na)\e{-2\pi \i n \zeta}.
\ee
In our case, we will need the case of $a=1$ which we consider from now on. The unitarity of ${\cal Z}$ can be easily shown, see for instance \cite{Neretin}. Let $f$ be a function in $L_1(\mathbbm{R})\cap L_2(\mathbbm{R})$ and we have 
\begin{equation*}
\int \limits_{\mathbbm{R}} \d x |f(x)|^2 =\frac{1}{2\pi}\infsum{k}\,\int\limits_{2\pi k+[0,2\pi]}\!\!\!\!\!\!\d x|f(x)|^2
=\frac{1}{2\pi}\int\limits_0^{2\pi}\d x\infsum{k}|f(x+2\pi k)|^2 <\infty,
\end{equation*}
where we used a partition of $\mathbbm{R}$ into intervals of $[0,2\pi]$ in the first step, the Fubini theorem in the second step and that $f$ is an $L_2$ function in the last step.  From this, we can conclude that $\infsum{k}|f(x+2\pi k)|^2 <\infty$ for a.e. $x\in\mathbbm{R}$.  Moreover, for $ f\in L_1(\mathbbm{R})\cap L_2(\mathbbm{R})$ the Fourier series $\infsum{k}f(x+2\pi k)\es{-2\i\pi k\zeta}$ is well-defined. Furthermore, we have
\begin{equation*}
\int\limits_0^1 \d\zeta |{\cal Z}(x,\zeta)|^2=\infsum{k}|f(x+2\pi k)|^2.    
\end{equation*}
Applying additionally an integration over $x$ and using again Fubini's theorem, we finally obtain
\begin{align*}
\norm{{\cal Z}[f]}^2_{L_2(\mathbbm{R}^2/\mathbbm{Z}^2)} &=
\frac{1}{2\pi}\int\limits_0^{2\pi}\d x\int\limits_0^1 \d\zeta|{\cal Z}(x,\zeta)|^2 \\ 
&=\frac{1}{2\pi}\int\limits_0^{2\pi}\d x\infsum{k}|f(x+2\pi k)|^2=\int\limits_{\mathbbm{R}} \d x |f(x)|^2=\norm{f}^2_{L_2(\mathbbm{R})}. 
\end{align*}
Using that $L_1(\mathbbm{R})\cap L_2(\mathbbm{R})$ is dense in $L_2(\mathbbm{R})$, the unitarity is shown. 

As has for instance been shown in \cite{Neretin}, the Zak transform can be 
extended to distributions and one obtains a bijection between the dual of the 
Schwartz space ${\cal S}^\prime(\mathbbm{R})$ and ${\cal Q}^\prime (\mathbbm{R}^2)$, where the latter is the dual of the space of all smooth functions that satisfy the quasiperiodicity condition in (\ref{eq:QuasiPer}). In particular, this allows to apply the Zak transform also to delta functions and as a further step to Gaussian distributions. The resulting image under the Zak transformation are Theta distributions as we will discuss below. 
\\
\\
In the complexifier approach, coherent states are constructed as analytic continuations of the heat kernel, yielding directly the coherent states of (\ref{eq:CSQM}). The self-similar solution of the heat equation with distributional boundary conditions, the so-called heat kernel, has the form
\begin{equation}
\rho_t(x,y)=\Cn\frac{1}{\sqrt{t}}\es{-\frac{(x-y)^2}{4\kdiff t}},    
\end{equation}
where $\kdiff$ denotes the diffusion constant and $\Cn$ a normalisation constant, see also section \ref{sec:SolHeat} for more details on self similar solutions of the heat equation. Later, we will choose $\Cn$ in such a way that our normalisation of the coherent states agrees with the one used in \cite{GCS1,GCS2,GCS3,GCS4,Bahr:2006kb,Brunnemann1,Brunnemann2}, where also complexifier coherent states are used. In order to obtain a normalised heat kernel, the choice $\Cn=\frac{1}{\sqrt{4\kdiff\pi}}$ is needed. A solution of the heat equation with boundary data $f(x)$ is then given by
\begin{equation*}
u(x,t)=\int\limits_{-\infty}^\infty \d y\, \rho_t(x,y)f(y).    
\end{equation*}
Following the notation of \cite{Neretin}, we define the Gaussian integral operator $B_\rho\colon {\cal S}(\mathbbm{R})\to{\cal S}(\mathbbm{R})$ as
\be
B_{\rho_t}[f](x,t)\coloneqq \int\limits_{\mathbbm{R}}\d y\, \rho_t(x,y)f(y)
\ee
and hence we have $B_{\rho_t}[f](x,t)=u(x,t)$.

The non-normalised coherent states constructed from the analytic continuation of the heat kernel are then given by
\be
\Pqp(x)=\left[\rho_{\hbar}(x,y)\right]_{y\to q+\i p}
=C_{q,p,\hbar}\es{-\frac{(x-q)^2}{2\hbar}}\e{\frac{\i }{\hbar}px},
\ee
with $C_{q,p,\hbar}=\Cn\exp\left(\frac{1}{2\hbar}(-2\i qp+p^2)\right)$, where we chose $\kdiff=\frac{1}{2}, m=1,\omega=1$ so that $\sigma=\sqrt{\hbar}$. In order to apply the Zak transform to the heat kernel, we choose $a=1$ and apply formula $(\ref{eq:ZakDilat})$. As has been already shown in \cite{Neretin} in a more general context, the Zak transform of the Gaussian integral operator  $B_{\rho_t}$ reads
\be
{\cal Z}B_{\rho_t}{\cal Z}^{-1}\colon{\cal Q}(\mathbbm{R}^2) \to {\cal Q}(\mathbbm{R}^2),\quad
{\cal Z}B_{\rho_t}{\cal Z}^{-1}[g](x,\zeta_1)\coloneqq \int\limits_{0}^{2\pi}\int\limits_{0}^1 {\cal K}_{\rho_t}(x,\zeta_1,y,\zeta_2)g(y,\zeta_2) \d y \d\zeta_2.
\ee
The kernel ${\cal K}_{\rho_t}(x,\zeta_1,y,\zeta_2)$ is related to the Zak transform of the heat kernel via
\be
{\cal K}_{\rho_t}(x,\zeta_1,y,\zeta_2)={\cal Z}[\rho_t](x,\zeta_1,y,-\zeta_2),\quad (x,\zeta_1),(y,\zeta_2)\in\mathbbm{R}^2,
\ee
where the minus sign in front of $\zeta_2$ results from the fact that the kernel is a function of $(x-y)$, see \cite{Neretin} for the general proof. 
The explicit form of the kernel then has the following form
\begin{align}
{\cal K}_{\rho_t}(x,\zeta_1,y,\zeta_2) &=\Cn
\frac{1}{\sqrt{t}}\sum\limits_{n,m=-\infty}^\infty \e{-\frac{(x+2\pi n - y-2\pi m)^2}{4\kdiff t}}\e{-2\i \pi n\zeta_1}\e{2\i \pi m\zeta_2}\nonumber  \\
&=\Cn
\frac{1}{\sqrt{t}}\sum\limits_{n,m=-\infty}^\infty \e{-\frac{(x-y + 2\pi(n+m))^2}{4\kdiff t}}\e{-2\i \pi n\zeta_1}\e{-2\i \pi m\zeta_2}\nonumber .
\end{align}
\\
In order to rewrite this again in terms of theta functions, we introduce $\widetilde{n}\coloneqq n+m$ and $\widetilde{m}\coloneqq n-m$, which leads to
\begin{align}
\label{eq:ThetaKernel}
{\cal K}_{\rho_t}(x,&\zeta_1,y,\zeta_2) =
\frac{\Cn}{\sqrt{t}}\sum\limits_{\widetilde{n},\tilde{m}=-\infty}^\infty \e{\frac{-(x-y+2\pi\widetilde{n})^2}{4\kdiff t}}\e{-2\i \pi\zeta_1\frac{1}{2}(\widetilde{n}+\tilde{m})}\e{-2\i \pi \zeta_2\frac{1}{2}(\widetilde{n}-\widetilde{m})} \nonumber \\
&=
\frac{\Cn\es{-\frac{(x-y)^2}{4\kdiff t}}}{\sqrt{t}}\sum\limits_{\widetilde{n},\widetilde{m}=-\infty}^\infty
\e{-\frac{2\pi\widetilde{n}}{2\kdiff t}(x-y)}\e{-\frac{(2\pi)^2}{4\kdiff t}\tilde{n}^2}\e{-2\i \pi\widetilde{n}\frac{1}{2}(\zeta_1+\zeta_2)} \e{-2\i \pi\widetilde{m}\frac{1}{2}(\zeta_1-\zeta_2)} \nonumber \\
&=
\frac{\Cn\es{-\frac{(x-y)^2}{4\kdiff t}}}{\sqrt{t}}
\Theta\left(-\frac{\pi}{2}(\zeta_1+\zeta_2)+\frac{2\i\pi(x-y)}{4\kdiff t},\frac{\i\pi}{\kdiff t}\right)
\sum\limits_{\widetilde{m}=-\infty}^\infty \e{-2\i \pi \widetilde{m}\frac{1}{2}(\zeta_1-\zeta_2)}.
\end{align}
The last sum on the right hand side does not convergence, but can be understood as a sequence of delta functions
\begin{equation*}
\delta_n(\zeta_2-\zeta_1)\coloneqq \sum\limits_{\widetilde{m}=-n}^n \e{\i \widetilde{m}\pi(\zeta_2-\zeta_1) }    
\end{equation*}
with
\begin{equation*}
\delta(\zeta_2-\zeta_1)=\lim\limits_{n\to\infty} \delta_n(\zeta_1-\zeta_2)=\sum\limits_{\tilde{m}=-\infty}^\infty \e{\i \tilde{m}\pi(\zeta_2-\zeta_1)} .
\end{equation*}
The Gaussian integral operator ${\cal Z}B_\rho{\cal Z}^{-1}$ involving the theta kernel from (\ref{eq:ThetaKernel}) is then given by
\begin{align}
{\cal Z} & B_{\rho_t}{\cal Z}^{-1}[g](x,\zeta_1)
= \nonumber \\
&\lim\limits_{n\to\infty}\frac{\Cn}{\sqrt{t}}
\int\limits_{T^2} \d y \d\zeta_2\, \e{-\frac{(x-y)^2}{4\kdiff t}} \Theta\left(-\frac{\pi}{2}(\zeta_1+\zeta_2)+\frac{\i\pi(x-y)}{2\kdiff t},\frac{\i\pi}{4\kdiff t}\right)\delta_n(\zeta_2-\zeta_1)g(y,\zeta_2),
\end{align}
where we used in the last line the definition for sequences of elements of the dual space. Performing the integration with respect to $\zeta_2$ as well as the limit, we obtain
\begin{align}
\label{eq:SolZakHeat}
u(x,t,\zeta_1)&\coloneqq {\cal Z}B_{\rho_t}{\cal Z}^{-1}[g](x,\zeta_1)
\nonumber \\
&=\frac{\Cn}{\sqrt{t}}\int\limits_{T^2}\d y \es{-\frac{(x-y)^2}{4\kdiff t}}\Theta\left(-\pi\zeta_1+\frac{\i\pi(x-y)}{2\kdiff t},\frac{\i\pi}{\kdiff t}\right)g(y,\zeta_1).
\end{align}
Then, by construction, $u(x,t,\zeta_1)\coloneqq {\cal Z}B_{\rho_t}{\cal Z}^{-1}[g](x,\zeta_1)$ is a solution of the push forward of the heat equation under the Zak transformation with quasi periodic boundary data $u(x,0,\zeta_1)=\sum\limits_{n}\e{-2\i n\pi\zeta_1}\tilde{g}(x,\zeta_1)$, where we have absorbed the normalisation constant $\Cn$ into a redefinition of $g(x,\zeta)$ to get $\tilde{g}(x,\zeta)$.

This becomes even more obvious when we look at the exact form of the pushed forward heat equation. As can be easily shown \cite{Neretin} just by direct computation, a push forward of the differential $\frac{\partial}{\partial x^j}$ reads ${\cal Z}_*\frac{\partial}{\partial x^j}=\frac{\partial}{\partial x^j}$, where as before and in abuse of notation we also call the new coordinates $x^j$. Considering this, the push forward of the heat equation is given by
\begin{equation*}
\left( \frac{\partial}{\partial t}-\kdiff\frac{\partial^2}{\partial x^2}\right)u(x,t,\zeta_1)=0.
\end{equation*}
The pushed forward differential operator of the heat equation does not depend on the variable $\zeta_1$, and thus the function $u(x,t,\zeta_1)$ in (\ref{eq:SolZakHeat}) is a solution of the heat equation for each fixed value of $\zeta_1$. In particular, we can choose $\zeta_1=0$. This additional $\zeta_1$-parameter can be identified with the $\delta$-parameter in \cite{Kastrup:2005xb,Kowalski_1996,Bahr:2006kb} and the $k$-parameter in \cite{Olmo}.
\\
\\
The above result for the theta kernel is also consistent with the convolution property of the Zak transform that reads for all $f_1,f_2\in L_2(\mathbbm{R})$ with $f_1*f_2\in L_1(\mathbbm{R})$ 
\be
{\cal Z}[f_1*f_2]={\cal Z}[f_1] *_{y} {\cal Z}[f_2],
\ee
where 
\be
(f_1*f_2)(x)\coloneqq \int\limits_{\mathbbm{R}} \d y\, f_1(x-y)f_2(y),\quad 
{\cal Z}[f_1] *_{y} {\cal Z}[f_2](x,\zeta)\coloneqq  \int\limits_0^1 \d y {\cal Z}[f_1](x-y,\zeta){\cal Z}[f_2](y,\zeta),
\ee
as for instance proved in \cite{Janssen1988}. This consistency can be easily checked by choosing $f_1=\rho_t$ and $f_2=g$ and considering the fact that the Zak transform of $\rho_t$ is given by
\be
{\cal Z}[\rho_t](x-y,\zeta)=\frac{\Cn\e{-\frac{(x-y)^2}{4\kdiff t}}}{\sqrt{t}}\Theta\left(-\pi\zeta+\frac{\i\pi(x-y)}{2\kdiff t},\frac{\i\pi}{\kdiff t}\right).
\ee
In order to rewrite the Theta kernel in a more compact form, we can further use the scaling property of the Theta function
\be
\label{ScalingTheta}
\Theta(z, \tau)=(-\i \tau)^{-\frac{1}{2}}\exp\lr{\frac{z^2}{\i \pi\tau}}\Theta\left(\frac{z}{\tau},-\frac{1}{\tau}\right).
\ee
For our specific case, we choose $\tau=\frac{\i\pi}{\kdiff t}$ and $z=-\pi\zeta+\frac{\tau}{2}(x-y)$
and obtain
\begin{equation*}
\Theta\left(-\pi\zeta+\frac{\i\pi(x-y)}{2\kdiff t},\frac{\i\pi}{\kdiff t}\right)
=\Theta\left(-\pi\zeta +\frac{\tau}{2}(x-y),\frac{\i\pi}{\kdiff t}\right)
=\Theta(z,\tau).
\end{equation*}
Furthermore, we have  $z^2/(\i\pi\tau)=-\zeta^2\kdiff t+\frac{(x-y)^2}{4\kdiff t}+\i\zeta(x-y)$ so that we can rewrite ${\cal Z}[\rho_t]$ in the following form
\begin{eqnarray}
{\cal Z}[\rho_t](x-y,\zeta)&=&\frac{\Cn}{\sqrt{t}}\frac{\sqrt{\kdiff t}}{\sqrt{\pi}}\e{-\zeta^2\kdiff t}\e{\i\zeta(x-y)}\Theta\left(-\frac{\pi\zeta}{\tau}+\frac{1}{2}(x-y),-\frac{1}{\tau}\right)
 \nonumber\\
&=&
\Cn\frac{\sqrt{\kdiff}}{\sqrt{\pi}}\e{-\zeta^2\kdiff t}\e{\i\zeta(x-y)}
\Theta\left(\i\zeta\kdiff t +\frac{1}{2}(x-y),-\frac{\kdiff t}{\i\pi}\right) \nonumber\\
&=&
\frac{\Cn\sqrt{\kdiff}}{\sqrt{\pi}}\sum\limits_{n\in\mathbbm{Z}}\e{-(n+\zeta)^2\kdiff t}\e{\i(n+\zeta)(x-y)},
\end{eqnarray}
where we used the definition of the $\Theta$-function in the last line. 
\\
\\
Given the map of the heat kernel under the Zak transform, we can now use this result to construct coherent states on the circle. Similar to the ordinary quantum mechanical case, we choose $t$ as our classicality parameter for this purpose and consider the analytic continuation of the image of the heat kernel. We then obtain for the non-normalised coherent states
\begin{eqnarray*}
\Psi^{t}_{\theta_0,p}(\phi ;\zeta)&=&\left[{\cal Z}[\rho_t](\phi-y,\zeta)\right]_{y=\theta_0+\i p} \\
&=& \frac{\Cn\sqrt{\kdiff}}{\sqrt{\pi}}\e{-\zeta^2\kdiff t}\e{\i\zeta(\phi-(\theta_0+\i p))} \, \Theta\left(\i\zeta\kdiff t +\frac{1}{2}(x-(\theta_0+\i p)),-\frac{\kdiff t}{\i\pi}\right) \nonumber\\
&=&
\frac{\Cn\sqrt{\kdiff}}{\sqrt{\pi}}\sum\limits_{n\in\mathbbm{Z}}\e{-(n+\zeta)^2\kdiff t}\e{\i(n+\zeta)\phi}\e{-\i(n+\zeta)\theta_0}\e{(n+\zeta)p},
\end{eqnarray*}
where we denoted the point $x\in S^1$ by $\phi$ in order to emphasise that these are coherent states on the circle for each fixed value of $\zeta$. That this is indeed the case has been shown in \cite{Olmo}. For the choice of $\Cn=\sqrt{\frac{\pi}{\kdiff}}$ and a diffusion constant of $\kdiff=\frac{1}{2}$, hence $\Cn=\sqrt{2\pi}$, these states exactly agree with the U(1) complexifier coherent states used in \cite{Bahr:2006kb} given by
\be
\label{eq:ComplU(1)usual}
\Psi^t_{\theta_0,p}(\phi;\zeta)=\sum\limits_{n=-\infty}^\infty \e{-(n+\zeta)^2\frac{t}{2}}\e{(n+\zeta)p}\e{-\i(n+\zeta)\theta_0}\e{\i(n+\zeta)\phi},
\ee
where the $\delta$ in \cite{Bahr:2006kb} needs to be identified with our $\zeta$. For the special choice of $\zeta=0$, we differ from the U(1) coherent states in \cite{GCS1,GCS2,GCS3,GCS4,Towards1,Towards2} by a sign change in front of $\theta_0$ and $\phi$. This is due to the fact that the Poisson algebra is defined according to $\{p,q\}=1$ in \cite{GCS1,GCS2,GCS3,GCS4,Towards1,Towards2} and the complexified $q$ then becomes $z=q-\i p$. Since the contribution from $\theta_0$ is just a phase and the contribution of the integral involving $\phi$ as far as expectation values are considered is invariant under the change $\phi\to -\phi$, our final results can, however, still be compared to the results in \cite{GCS2,GCS3}. 

In the following, we will keep the variable $\zeta$ as an arbitrary parameter in the U(1) coherent states computations because this allows us to map straightforwardly between the U(1) case and the standard case in $L_2(\mathbbm{R})$ by means of the Zak transform, as we will present below. We moreover fix the diffusion constant $\kdiff=\frac{1}{2}$ from now on in order to compare our results more easily to \cite{GCS1,GCS2,GCS3,GCS4,Bahr:2006kb}.
\section{Semiclassical expectation values using Kummer functions}
In this section, we compute the semiclassical expectation values of the dynamical operators discussed already in the quantum mechanical case in section \ref{sec:QMCase} for coherent states on the circle. As shown in the last section, the non-normalised U(1) coherent states are then given by
\begin{align}
\label{eq:GenU(1)CS}
\Psi^{t}_{\theta_0,p}(\phi;\zeta)=
\sum\limits_{n=-\infty}^\infty 
\e{-\frac{t}{2}(n+\zeta)^2}\e{\i (n+\zeta)(\phi-(\theta_0+\i p))}.
\end{align}

\subsection{Computation of expectation values of \texorpdfstring{$\lrabs{\hat p}^r$}{pr}}
The corresponding momentum operator of U(1) acts on these coherent states in the following way:
\be
\hat{p}\Psi^{t}_{\theta_0,p}(\phi;\zeta)=-\i t \frac{\d}{\d\phi} \Psi^{t}_{\theta_0,p}(\phi;\zeta)=
\sum\limits_{n=-\infty}^\infty (n+\zeta)t\es{-\frac{t}{2}(n+\zeta)^2}\e{\i (n+\zeta)(\phi-(\theta_0+\i p))} \label{eq:pAction}.
\ee
The expectation value that we are interested in then reads
\be
\langle \Psi^{t}_{\theta_0,p}(\zeta)\,|\, |\hat{p}|^r \, |\,  \Psi^{t}_{\theta_0,p}(\zeta)\rangle
=\norm{\Psi^{t}_{\theta_0,p}}^{-2}\sum\limits_{n\in\mathbbm{Z}}
|(n+\zeta)t|^r \e{-t(n+\zeta)^2}\e{2(n+\zeta)p}.
\ee
Compared to the usual case where $\zeta$ is fixed to be zero, we realise that the way the non-vanishing $\zeta$ enters into the expectation value is as a kind of shift of $n$. Therefore, if we apply the Poisson summation formula now and use the modulation property of the Fourier transform,  the result is the same as in the standard case up to an additional exponential of the form $\e{2\i N\pi\zeta}$, where $N$ denotes the summation index after the Poisson summation has been performed. Thus, we obtain
\begin{align}
\langle \Psi^{t}_{\theta_0,p}(\zeta)\,|\, |\hat{p}|^r\, |\,  \Psi^{t}_{\theta_0,p}(\zeta)\rangle =\norm{\Ptg(\zeta)}^{-2}\frac{\sqrt{2\pi}}{T}T^r\infsum{N} \e{2\i \pi N\zeta}\infintpi{x}|x|^r \e{-x^2+\frac{2p}{T}x-\frac{2\pi\i N}{T}x},
\end{align}
where we defined $x\coloneqq n+\zeta, T\coloneqq \sqrt{t}$ and used the translation invariance of $\d x$. Now, we can proceed as in the quantum mechanical case. We rewrite $|x|^r$ in terms of the Kummer function of the second kind and afterwards write the integrand above as a product of the Fourier transform of the Kummer function, a Gaussian and a complex exponential. This yields
\begin{flalign}
&\langle \Psi^{t}_{\theta_0,p}(\zeta)\,|\, |\hat{p}|^r\, |\,  \Psi^{t}_{\theta_0,p}(\zeta)\rangle && \nonumber\\
&=\norm{\Ptg(\zeta)}^{-2}\frac{\sqrt{2\pi}}{T}T^r\infsum{N}
\e{2\i \pi N\zeta}\infintpi{x}
U\left(-\frac{r}{2},-\frac{r}{2}+1,x^2\right)\e{-x^2}\e{\i x\left(\frac{-2\i p}{T}\right)}\e{-\frac{2\pi\i N}{T}x} && \nonumber \\
&= \norm{\Ptg(\zeta)}^{-2}\frac{\sqrt{2\pi}}{T}T^r\infsum{N}
\e{2\i \pi N\zeta}
{\cal F}\left(U\left(-\frac{r}{2},-\frac{r}{2}+1,x^2\right)\e{-x^2}\e{\i x\left(\frac{-2\i p}{T}\right)}\right)\left(\tfrac{2\pi N}{T}\right) &&\nonumber \\
&= \norm{\Ptg(\zeta)}^{-2} \frac{\sqrt{2\pi}}{T}T^r\frac{1}{\sqrt{2}} \frac{\Gamma\left(\frac{r+1}{2}\right)}{\Gamma\left(\frac{1}{2}\right)}
\infsum{N}\e{2\i \pi N\zeta}
\e{-\frac{\big(\frac{2\pi N}{T}+\frac{2\i p}{T}\big)^2}{4}} \kchf{-\frac{r}{2},\frac{1}{2},\frac{\big(\frac{2\pi N}{T}+\frac{2\i p}{T}\big)^2}{4}}
&& \nonumber \\ 
&= \norm{\Ptg(\zeta)}^{-2} \frac{T^r}{T} \Gamma\left(\tfrac{r+1}{2}\right)
\infsum{N}\e{2\i \pi N\zeta}
\e{-\frac{(\pi N+\i p)^2}{T^2}}\kchf{-\frac{r}{2},\frac{1}{2},\left(\frac{\pi N}{T}+\frac{\i p}{T}\right)^2} \label{eq:ExpprZeta}
&&
\end{flalign}
where we applied lemma \ref{LemModFourier} in the second step and also used $\Gamma\lr{\frac{1}{2}}=\sqrt{\pi}$. Using the Kummer transformation for $\kchf{a,b,z}$ shown in (\ref{eq:KummerTrafo}), we finally obtain
\begin{flalign}
\label{eq:ExpFinZeta}
&\langle \Psi^{t}_{\theta_0,p}(\zeta)\,|\, |\hat{p}|^r\, |\,  \Psi^{t}_{\theta_0,p}(\zeta)\rangle && \nonumber\\
&=\norm{\Ptg(\zeta)}^{-2} \frac{T^r}{T} \Gamma\left(\tfrac{r+1}{2}\right)
\infsum{N}\e{2\i \pi N\zeta}
\kchf{\frac{r+1}{2},\frac{1}{2},\left(\frac{p-\pi\i N}{T}\right)^2}.
\end{flalign}
To complete this calculation, let us compute the norm of the coherent states that is involved in the result above. We get after the Poisson summation
\be
\label{eq:normZeta}
\norm{\Ptg(\zeta)}^2
=\infsum{n} \e{-t(n+\zeta)^2}\e{2(n+\zeta)p}
=\sqrt{\frac{\pi}{T^2}}\infsum{N} \e{2\pi\i N\zeta}
\es{-\frac{(\i \pi p+N\pi)^2}{T^2}}.
\ee

Being now interested in the asymptotics of $t$ being small, and hence $T$ accordingly, we can apply the asymptotic expansion for large arguments of the KCHF,~\eqref{expansion}, on~\eqref{eq:ExpFinZeta}. This follows closely the procedure of the quantum mechanical scenario of~\eqref{eq:QMprExpansion}. With the norm~\eqref{eq:normZeta}, a Gaussian in $\nicefrac{p^2}{t}$ enters the expectation value. Hence, out of the two series of the KCHF's asymptotic expansion, only the one with the inverse Gaussian in $\nicefrac{p^2}{t}$ remains, while the other one is damped to $\mathcal{O}\lr{t^{\infty}}$. In the end, we obtain
\begin{align}
    \langle \Psi^{t}_{\theta_0,p}(\zeta)\,|\, |\hat{p}|^r\, |\,  \Psi^{t}_{\theta_0,p}(\zeta)\rangle &= \lrabs{p}^r \sum_{n=0}^{\infty} \frac{\lr{-\frac{r}{2}}_n \lr{\frac{1-r}{2}}_n}{n!}\lr{\frac{t}{p^2}}^{\!n} \nonumber\\
    &= \lrabs{p}^r \lr{1-\frac{r\lr{1-r}}{4}\frac{t}{p^2}+\mathcal{O}\lr{t^2}},
\end{align}
which resembles perfectly the quantum mechanical result of~\eqref{eq:QMprExpansion}. That this is expected will be discussed in detail in section \ref{sec:RelQMU1}, where the relation between semiclassical expectation values of quantum mechanics and U(1) is analysed by means of the Zak transform and its properties. 

\subsection{Computation of expectation values of \texorpdfstring{${\hat q}$}{q}}
\label{sec:qU1}
As introduced in section~\ref{sec:QMCase}, the operators ${\hat q}(r)$ are of importance in loop quantum gravity. With the previously obtained results of~\eqref{eq:ExpFinZeta} and~\eqref{eq:normZeta}, we may now also rigorously compute expectation values of them in U(1). From~\eqref{eq:ExpFinZeta}, we can directly derive the expectation value of $\lrabs{\hat p}^r$ w.r.t. the coherent state the inverse holonomy acted on:
\begin{align}
    &\langle {\hat h}\inv \Psi^{t}_{\theta_0,p}(\zeta)\,|\, |\hat{p}|^r\, |\,  {\hat h}\inv\Psi^{t}_{\theta_0,p}(\zeta)\rangle && \nonumber\\
&=\norm{\Ptg(\zeta)}^{-2}\frac{T^r}{T} \Gamma\left(\tfrac{r+1}{2}\right)
\infsum{N}\e{2\i \pi N\zeta}
\kchf{\frac{r+1}{2},\frac{1}{2},\left(\frac{p-T^2-\pi\i N}{T}\right)^{\!2}}.
\end{align}
Like in the standard quantum mechanical case of section~\ref{sec:QMCase}, the inverse holonomy acting on the coherent state causes an (infinitesimal) shift in the momentum. With the holonomy being $h=\e{\i\phi}$, the momentum operator's $\phi$-derivative now not only sees $\e{-\i (n+\zeta)\phi}$ as in~\eqref{eq:pAction}, but acts on $\e{\i (n+\zeta)\phi}\e{-\i\phi} = \e{\i (n+\zeta-1)\phi}$ instead. Hence, $\lrabs{\hat p}^r$ acting on the shifted coherent state evaluates now $\lrabs{\lr{n+\zeta-1}t}^r$ and we can cast that shift into $p\mapsto p-t$ in the state's exponentials via a redefinition of $n\mapsto n-1$.

Having these two expectation values of $\lrabs{\hat p}^r$, we can combine them to the commutator's expectation value and proceed as before in the quantum mechanical case, namely by performing the asymptotic expansion for large arguments of the KCHFs, as both of them grow with $\frac{1}{t}$ for small $t$. In the end, we obtain
\begin{align}
    \langle {\hat q}^r \rangle_{\Psi^t_{\theta_0,p}} \approx \frac{r\lrabs{p}^r}{p}t + \lrabs{p}^r\lr{\frac{r\lr{1-r}}{2p^2} + \frac{r\lr{2-3r+r^2}}{4p^3} }t^2 + \mathcal{O}\lr{t^3}.
\end{align}

We see again that the series' first term corresponds one-to-one to the derivative's result. For the next order, we got two contributions: one that resembles the second derivative but also a further one. If we compare the result above to the quantum mechanic's result of~\eqref{eq:QMdirectKCHF}, we notice that the $\hbar^2$-contribution there also comprised two terms. While one included $\alpha^2$, namely the one with the numerical prefactors and powers of $p$ in accordance to the second derivative, the second one was proportional to $\alpha$.
\section{Relation of semiclassical matrix elements for \texorpdfstring{$L_2(S_1)$}{L2S1} and \texorpdfstring{$L_2(\mathbbm{R})$}{L2R} coherent states}
\label{sec:RelQMU1}

As discussed in section \ref{sec:ZakTrafo}, the Zak transformation provides a unitary map between the Hilbert spaces $L_2(\mathbbm{R})$ and $L_2(\mathbbm{R}^2/\mathbbm{Z}^2)$. Therefore, the expectation values of the usual quantum mechanical operators and their corresponding Zak-transformed counterparts are identical.
However, if we compute matrix elements or semiclassical expectation values with respect to U(1) coherent states for fixed $\zeta$, as one does for coherent states on a circle, we only perform one of the integrals, namely the one over $\theta$, out of the two integrations involved in the inner product of $L_2(\mathbbm{R}^2/\mathbbm{Z}^2)$. Nevertheless, as we will show below, the expectation value with respect to U$(1)$ coherent states is completely determined if we know the matrix elements of the corresponding operator with respect to the harmonic oscillator coherent states in $L_2(\mathbbm{R})$. As shown above, the U(1) coherent states can be written by means of the Zak transform as 
$ \Ptg(\phi,\zeta)={\cal Z}[\Pqp](\phi,\zeta)$, and accordingly operators $\hat{O}_{\rm QM}$ transform under ${\cal Z}$ as ${\cal Z}\hat{O}_{\rm QM}{\cal Z}^{-1}$. Hence, the integrand of an $L_2(S_1)$ expectation value with respect to U(1) coherent states is given by
\begin{equation*}
\overline{{\cal Z}[\Pqp]}(\phi,\zeta){\cal Z}\hat{O}_{\rm QM}{\cal Z}^{-1}{\cal Z}[\Pqp](\phi,\zeta)
=\overline{{\cal Z}[\Pqp]}(\phi,\zeta){\cal Z}[\hat{O}_{\rm QM}\Pqp](\phi,\zeta).
\end{equation*}

In the following considerations, we want to examine the relation between $\int_0^{2\pi}\frac{\d\phi}{2\pi}\overline{{\cal Z}[\Pqp]}(\phi,\zeta){\cal Z}[\hat{O}_{\rm QM}\Pqp](\phi,\zeta)$ and matrix elmenents $\langle\Psi^\hbar_{q',p'}\,|\, \hat{O}_{\rm QM}\,|\, \Pqp\rangle_{L_2(\mathbbm{R})}$. We will follow \cite{Janssen1988}, where part of the formulas are presented but partly without proofs\footnote{Note that in \cite{Janssen1988}, some definitions might differ by factors of $\pi$ because we adopted the operators to the case needed for our work.}. With $f,g \in L_2(R)$, obviously $\overline{{\cal Z}[f]}(\phi,\zeta){\cal Z}[g](\phi,\zeta)$ is periodic in $\zeta$, which follows directly from the definition of the Zak transform. Furthermore, it is also $2\pi$-periodic in $\phi$. We have
\begin{align*}
\overline{{\cal Z}[f]}(\phi+2\pi,\zeta){\cal Z}[g](\phi+2\pi,\zeta)&=
\infsum{m,n}\overline{f}(\phi+2\pi+2\pi m)\es{\i2\pi m\zeta}g(\phi+2\pi+2\pi n)\es{-\i2\pi n\zeta} \\
&\hspace{-.4cm}\stackrel{\substack{r\coloneqq m+1 \\ s\coloneqq n+1}}{=}\infsum{r,s}\overline{f}(\phi+2\pi r)\es{\i2\pi(r-1)\zeta}g(\phi+2\pi s)\es{-\i2\pi(s-1)\zeta} \\
&=
\infsum{r,s}\overline{f}(\phi+2\pi r)\es{\i 2\pi r\zeta}g(\phi+2\pi s)\es{-\i 2\pi s\zeta} \\
&=
\overline{{\cal Z}[f]}(\phi,\zeta)\;\!{\cal Z}[g](\phi,\zeta).
\end{align*}
As a consequence, we can expand ${\cal Z}[f]{\cal Z}[g]$ into a Fourier series given by
\begin{equation}
(\overline{{\cal Z}[f]}{\cal Z}[g])(\phi,\zeta)=\infsum{m,n}{\cal F}_{mn}\es{\i xm}\es{2\pi\i n\zeta},    
\end{equation}
where ${\cal F}_{mn}$ denote the Fourier coefficients that have the form
\begin{equation}
\label{eq:Fmn}
{\cal F}_{mn}=\int\limits_0^{2\pi}\frac{\d\phi}{2\pi}\int\limits_0^1 \d\zeta  
\left(\overline{{\cal Z}[f]}{\cal Z}[g]\right)(\phi,\zeta)\es{-\i mx}\es{-2\pi\i n\zeta}.
\end{equation}
Next, we introduce the translation and scaling operators defined by
\be
(T_af)(x)=f(x+a)\quad\text{and}\quad (R_bf)(x)=\e{-\i bx}f(x), \quad a,b\in\mathbbm{R}.
\ee
The Fourier coefficients ${\cal F}_{mn}$ can be easily computed as already discussed in \cite{Janssen1988} and in more detail in \cite{Zayed1995}:
\begin{lemma}
The Fourier coefficients ${\cal F}_{mn}$ in 
\be
(\overline{{\cal Z}[f]}{\cal Z}[g])(\phi,\zeta)=\infsum{m,n}{\cal F}_{mn}\es{\i xm}\es{2\pi\i n\zeta}   
\ee
are given by 
\be
{\cal F}_{mn}=\langle R_{-m}T_{2\pi n}f\;\! ,\;\! g\rangle_{L_2(\mathbbm{R})}.
\ee
\end{lemma}
In order to proof the lemma above, we just have to compute the Zak transform of $R_{-m}T_{2\pi n} f$. We obtain
\begin{eqnarray}
{\cal Z}[R_{-m}T_{2\pi n}f](\phi,\zeta) 
&=& 
\infsum{k}(R_{-m}T_{2\pi n}f)(\phi+2\pi k)\es{-2\i\pi k\zeta}\nonumber \\
&=&\infsum{k}f(\phi+2\pi(k+n))\es{-2\i\pi k\zeta}\es{\i(\phi+2\pi(k+n))m}\nonumber \\
&=&\infsum{k}f(\phi+2\pi k)\es{2\i\pi n\zeta}\es{-2\i\pi k\zeta}\es{\i\phi m} \nonumber \\
&=&{\cal Z}[f](\phi,\zeta)\es{\i\phi m}\es{2\pi\i\zeta n},
\end{eqnarray}
where we used the quasi periodicity of ${\cal Z}$ in the second last step. Given this, we obtain for the complex conjugate 
\be
\overline{{\cal Z}[R_{-m}T_{2\pi n}f]}(\phi,\zeta)
=\overline{{\cal Z}[f]}(\phi,\zeta)\e{-\i\phi m}\e{-2\pi\i\zeta n}.
\ee
Reinserting this back into the Fourier coefficients ${\cal F}_{mn}$, we get
\begin{eqnarray*}
{\cal F}_{mn} &=&
\int\limits_0^{2\pi}\frac{\d\phi}{2\pi}\int\limits_0^1 \d\zeta \overline{{\cal Z}[R_{-m}T_{2\pi n}f]}{\cal Z}[g](\phi,\zeta)\\
&=&
\langle \overline{{\cal Z}[R_{-m}T_{2\pi n}f]}\;\! ,\;\! {\cal Z}[g]\rangle_{L_2(S_1\times S_1^*)} \\
&=&\langle R_{-m}T_{2\pi n}f\;\! ,\;\! g\rangle_{L_2(\mathbbm{R})},
\end{eqnarray*}
where we used the unitarity of the Zak transform in the last step. Therefore, the Fourier series associated with $\langle{\overline{\cal Z}[f]}(\zeta)\;\! ,\;\! {\cal Z}[g](\zeta)\rangle_{L_2(S_1)}$ reads
\begin{eqnarray}
\label{eq:RelQMU1form1}
\langle{\overline{\cal Z}[f]}(\zeta)\;\! ,\;\! {\cal Z}[g](\zeta)\rangle_{L_2(S_1)}
&=&
\int\limits_0^{2\pi}\frac{\d\phi}{2\pi}\infsum{m,n}\langle R_{-m}T_{2\pi n}f\;\! ,\;\! g\rangle_{L_2(\mathbbm{R})} \es{\i m\phi}\es{2\i\pi n\zeta} \nonumber \\
&=&  \infsum{n}\langle T_{2\pi n}f\;\! ,\;\! g\rangle_{L_2(\mathbbm{R})}\es{2\i\pi n\zeta}.
\end{eqnarray}
If we apply the result in (\ref{eq:RelQMU1form1}) on the harmonic oscillator coherent states, we obtain the following relation between semiclassical matrix elements in $L_2(S_1)$ and $L_2(\mathbbm{R})$:
\begin{lemma}
\label{lem:RelQMU1}
For a linear operator $\hat{O}$ on (a suitable domain of) $L_2(S_1)$ that is obtained from the corresponding operator $\hat{O}_{\rm QM}$ on (a suitable domain of) $L_2(\mathbbm{R})$ by $\hat{O}={\cal Z}\hat{O}_{\rm QM}{\cal Z}^{-1}$ we have the following relation between the matrix elements in $L_2(S_1)$ and $L_2(\mathbbm{R})$:
\begin{equation}
\label{eq:FinRelQMU1}
\langle\Psi^t_{\theta_0^\prime,p'}(\zeta)\,|\,\hat{O}\, |\,\Ptg(\zeta)\rangle_{L_2(S_1)}
=
\infsum{n}\e{2\i\pi n\zeta}
\langle T_{2\pi n}\Psi^\hbar_{q',p'}\, |\, \hat{O}_{\rm QM}\, |\Pqp\rangle_{L_2(\mathbbm{R})}\big|_{\substack{\hspace{-1.0cm}t=\hbar \\ \theta_0=q\bmod{2\pi}\\ \theta^\prime_0=q'\bmod{2\pi}}} .
\end{equation}
\end{lemma}
We realise the semiclassical matrix elements in $L_2(S_1)$ can be understood as a Fourier series in $\zeta$ with Fourier coefficients $c_n\coloneqq \langle T_{2\pi n}f,g\rangle_{L_2(\mathbbm{R})}$. This means that the matrix elements in $L_2(S_1)$ are completely determined by the corresponding matrix elements in the ordinary quantum mechanics case in $L_2(\mathbbm{R})$. 
In particular, an additional integration over $\int_0^1d\zeta$ just projects onto the $n=0$ coefficient and this yields, as expected from the unitarity of ${\cal Z}$, the  semiclassical matrix element in $L_2(\mathbbm{R})$. 

Expectation values of U(1) coherent states for integer powers of holonomy and momentum operators have been computed in \cite{GCS2,GCS3,Bahr:2006kb} using complexifier coherent states. In \cite{GCS2,GCS3}, these were computed for the special case of $\zeta=0$, whereas in \cite{Bahr:2006kb} an arbitrary $\zeta$ was considered. In both works, they show that it is justified to only consider the $n=0$ term if we are interested in the classical limit $t\to 0$. Given this, we know from the lemma above that in this limit the result for $L_2(S_1)$ and $L_2(\mathbbm{R})$ exactly coincide and the value of $\zeta$ is irrelevant. In \cite{Olmo,Kastrup:2005xb,Kowalski_1996}, these kind of expectation values were computed and higher $n$ terms were rewrittten in terms of Jacobi's theta function and its derivative respectively. We believe that the relation shown in the lemma above yields a simpler formulation of the higher order terms in $n$ and easily allows to extract the semiclassical limit $t\to 0$.
\subsection{Application of lemma~\ref{lem:RelQMU1}}
To demonstrate that lemma~\ref{lem:RelQMU1} from the last section might simplify computations of semiclassical matrix elements and expectations values, we apply it to a couple of examples. First, we consider the overlap of two coherent states and thus choose $f=\Psi^\hbar_{q',p'}$ and $g=\Pqp$. The matrix element of interest in $L_2(\mathbbm{R})$ is then
\begin{equation}
\label{eq:OverlapL2R}
\langle T_{2\pi n}\Psi^\hbar_{q',p'}\,| \Pqp\rangle_{L_2(\mathbbm{R})}
=\sqrt{\frac{\pi}{\hbar}}\es{-\frac{1}{\hbar}\left(\frac{q-q'+\i(p+p')}{2}+n\pi\right)^2}.
\end{equation}
Given this, the corresponding overlap  for $L_2(S_1)$ reads
\begin{equation}
\label{eq:OverlapS1}
\langle\Psi^t_{\theta_0',p'}(\zeta)\, |\, \Ptg(\zeta)\rangle_{L_2(S_1)}
=\sqrt{\frac{\pi}{t}}\infsum{n}\e{-\frac{1}{\hbar}\left(\frac{q-q'+\i(p+p')}{2}+n\pi\right)^2}\e{2\pi\i n\zeta},
\end{equation}
which exactly agrees with the result in \cite{Bahr:2006kb} and reproduces also the correct result for the norm, for instance computed in \cite{GCS1,GCS3,Bahr:2006kb}, as a special case:
\begin{equation*}
\langle\Ptg(\zeta)\, |\, \Ptg(\zeta)\rangle_{L_2(S_1)}=\norm{\Ptg}^2
=\sqrt{\frac{\pi}{t}}\infsum{n}\e{-\frac{1}{t}(\i p+n\pi)^2}\e{2\pi\i n\zeta}.
\end{equation*}
In our further discussion, we will consider semiclassical expectation values of the basic operators as well as matrix elements for fractional powers of the momentum operator. We start with integer powers of holonomy operators, that is $\hat{h}^m=\e{\i m\hat{x}}$. For this operator, we get
\begin{equation*}
\langle T_{2\pi n}\Pqp\,|\,\e{\i m\hat{x}}\,|\,\Pqp\rangle_{L_2(\mathbbm{R})}
=\sqrt{\frac{\pi}{\hbar}}\es{-\frac{1}{\hbar}(\i p+n\pi)^2}\e{\i m(q-\frac{m\hbar}{4})}\e{-\i\pi nm},
\end{equation*}
and therefore the corresponding semiclassical expectation value for U(1) coherent states yields
\begin{equation*}
\langle\Ptg(\zeta)\, |\, \e{\i m\hat{\phi}}\,|\, \Ptg(\zeta)\rangle_{L_2(S_1)}
=\sqrt{\frac{\pi}{t}}\es{\i m(q-\frac{mt}{4})}\infsum{n}\e{-\frac{1}{t}(\i p+n\pi)^2}\e{-\i\pi nm}\e{2\pi\i n\zeta},
\end{equation*}
which agrees for $m=1$ and up to a different normalisation with the results in \cite{Kastrup:2005xb,Kowalski_1996}. We cannot directly compare it to the results in \cite{GCS3}, since they only considered the limit $t\to 0$ and thus some terms were neglected during the computation. 
\\
Let us briefly discuss the matrix elements in $L_2(\mathbbm{R})$ that enter the Fourier expansion of semiclassical expectation values in $L_2(S_1)$. For an operator of the form $\hat{O}=f(\hat{x})$, the Fourier coefficients have the following form:
\begin{equation*}
\langle T_{2\pi n}\Pqp\,|\, f(\hat{x})\,|\,\Pqp\rangle_{L_2(\mathbbm{R})}
=\frac{1}{\hbar}\es{-\frac{1}{t}(\i p+n\pi)^2}\int\limits_{\mathbbm{R}}\d x f(x)\es{-\frac{1}{\hbar}(x-q+\pi n)^2}.
\end{equation*}
We realise that the only difference to the case $n=0$, when the translation operator becomes the identity operator, is that (some) $q$ and $p$ labels get shifted by $n\pi$ or $-n\pi$ respectively. Note that this cannot be carried over to a shift in the $q,p$ labels for the entire state $\Pqp$ since also the normalisation constant $C_{q,p,\hbar}$ depends on these labels and no shift occurs there. Likewise, considering the Fourier transform of the states $\Pqp$, we can write down a smilar result for operators $\hat{O}=f(\hat{p})$ given by
\begin{equation*}
\langle T_{2\pi n}\Pqp\,|\, f(\hat{p})\,|\,\Pqp\rangle_{L_2(\mathbb{R})}
=\frac{1}{\hbar}\int\limits_{\mathbbm{R}}\d k \es{-\frac{1}{\hbar}k^2}\e{\frac{\i}{\hbar}k(-\i p+n\pi)}f(k).
\end{equation*}
Now, if we choose as an example $f(\hat{p})=|\hat{p}|^r$, we can easily show that the result in (\ref{eq:ExpFinZeta}) is consistent with lemma~\ref{lem:RelQMU1}, which also explains the additional shift by $n\pi$ in the argument of the Kummer function compared to the result for the quantum mechanical expectation value in $L_2(\mathbbm{R})$ of (\ref{eq:KCHFpr}) taking into account that $T=\sqrt{t}$ yielding a consistency check of our computations in the former sections. Moreover, for $n=0$ the results of (\ref{eq:ExpprZeta}) and (\ref{eq:KCHFpr}) agree as required if the normalisation is taken correctly.

Finally, we present the computation of matrix elements for fractional powers of the momentum operator, that is we are aiming at computing $\langle\Psi^t_{\theta_0',p'}(\zeta)\, |\, |\hat{p}|^r\,|\, \Ptg(\zeta)\rangle_{L_2(S_1)}$ by applying lemma \ref{lem:RelQMU1}.  For this purpose we need the explicit form of 
\begin{equation*}
\langle T_{2\pi n}\Psi^{\hbar}_{q'p'}\,|\, |\hat{p}|^r\,|\,\Pqp\rangle_{L_2(\mathbb{R})}    
=\int \d k\, {\cal F}(\overline{\Psi^{\hbar}_{q'p'})}(k)|k|^r{\cal F}(\Pqp)(k)e^{-\frac{\i}{\hbar}2\pi n k}.
\end{equation*}
Reinserting the Fourier transform ${\cal F}(\Pqp)(k)=C_{q,p,\hbar}e^{-\frac{1}{2\hbar}(k-p)^2}e^{-\frac{\i}{\hbar}(k-p)q}$ as well as that $C_{q,p,\hbar}e^{-\frac{1}{2\hbar}p^2+\frac{\i}{\hbar}pq}=\frac{1}{\sqrt{\hbar}}$ and $\overline{C}_{q',p',\hbar}e^{-\frac{1}{2\hbar}(p')^2-\frac{\i}{\hbar}p'q'}=\frac{1}{\sqrt{\hbar}}$ we obtain
\begin{eqnarray*}
\langle T_{2\pi n}\Psi^{\hbar}_{q'p'}\,|\, |\hat{p}|^r\,|\,\Pqp\rangle_{L_2(\mathbb{R})}    
&=&
\hbar^{\frac{r}{2}-1}\int \d k\, \lrabs{\frac{k}{\sqrt{\hbar}}}^r e^{-\frac{1}{\hbar}k^2}
e^{\frac{\i}{\sqrt{\hbar}}k\left(\frac{q'-\i p'}{\sqrt{\hbar}} - \frac{q+\i p}{\sqrt{\hbar}} - \frac{2\pi n}{\sqrt{\hbar}}\right)}.
\end{eqnarray*}
Next, we rewrite the absolute value again in terms of the Kummer function of the first kind, that is $\Big|\frac{k}{\sqrt{\hbar}}\Big|^r=U(-\frac{r}{2}, -\frac{r}{2}+1,\frac{k^2}{\hbar})$. Further, we introduce the new variable $\tilde{k}\coloneqq\frac{k}{\sqrt{\hbar}}$ and perform the integration by using the duality of Kummer's function of the first and second kind under Fourier transformations discussed in \ref{sec:IntroKummer}. This yields
\begin{align}
\label{eq:MEprL2R}
\langle T_{2\pi n}\Psi^{\hbar}_{q'p'} \,|\, & |\hat{p}|^r\,|\,\Pqp\rangle_{L_2(\mathbb{R})}=  \\   
&=
\sqrt{\frac{\pi}{\hbar}}\frac{\Gamma(\frac{r+1}{2})}{\Gamma(\frac{1}{2})}\hbar^{\frac{r}{2}}
e^{-\frac{1}{\hbar}\left(\frac{q-q'+\i(p+p')}{2}+n\pi\right)^2}
\kchf{-\tfrac{r}{2},\tfrac{1}{2},\tfrac{1}{\hbar}\left(\tfrac{q-q'+\i(p+p')}{2}+n\pi\right)^2}\nonumber .
\end{align}
If we specialise the above result to the case $r=0$ and consider that $\kchf{0,\frac{1}{2},z}=1$ for all $z$, then the result exactly agrees with the overlap shown in \eqref{eq:OverlapL2R}. Given the result for the matrix element in $L_2(\mathbb{R})$ by means of lemma \ref{lem:RelQMU1} we immediately get the result for $L_2(S_1)$, which finally takes the form
\begin{align}
\label{eq:ResMEprS1}
 \langle\Psi^t_{\theta_0',p'}&(\zeta)\, |\, |\hat{p}|^r\,|\, \Ptg(\zeta)\rangle_{L_2(S_1)} =\\
&=
\sqrt{\frac{\pi}{t}}\frac{\Gamma(\frac{r+1}{2})}{\Gamma(\frac{1}{2})}t^{\frac{r}{2}}
\sum\limits_{n=0}^\infty e^{\i2\pi n\zeta}e^{-\frac{1}{t}\left(\frac{q-q'+i(p+p')}{2}+n\pi\right)^2}
\kchf{-\tfrac{r}{2},\tfrac{1}{2},\tfrac{1}{t}\left(\tfrac{q-q'+\i(p+p')}{2}+n\pi\right)^2}\nonumber.
\end{align}
Comparing the final result in \eqref{eq:ResMEprS1} to the result in \eqref{eq:ExpprZeta} we realise that if we choose $\theta_0'=\theta_0$ and $p'=p$ corresponding to $q'=q, p'=p$ in \eqref{eq:ResMEprS1} as well as take into account that $T=\sqrt{t}$, then for this special case the results in \eqref{eq:ExpprZeta} and \eqref{eq:ResMEprS1} exactly coincide. As can be seen for all examples discussed in this subsection, lemma \ref{lem:RelQMU1} provides a method to compute semiclassical matrix elements for $L_2(S_1)$ by computing the corresponding shifted matrix elements in $L_2(\mathbb{R})$. In the next subsection we will discuss the relation between the Zak transform and the Poisson summation formula.
\subsection{The Zak transformation and the Poisson summation formula}
In the framework of complexifier coherent states, the Poisson summation is heavily used in computations of semiclassical matrix elements. For the benefit of the reader, we therefore review the relation between the Zak transformation and the Poisson summation here by following \cite{Wilson:2011}. In order to relate the Zak transformation to the Poisson summation formula, one additionally introduces a dual Zak transformation $\widetilde{\cal Z}$ defined as
\begin{equation*}
\widetilde{\cal Z}[f](x,\zeta)=\infsum{n}f(\zeta+n)\es{\i n x}    .
\end{equation*}
For  $g(x,\zeta)={\cal Z}[f](x,\zeta)$ and $\tilde{g}(x,\zeta)=\widetilde{\cal Z}[f](x,\zeta)$, the inverse of ${\cal Z}$ and $\widetilde{\cal Z}$ are given by
\begin{equation*}
{\cal Z}^{-1}[g](x)=\int\limits_0^1 \d\zeta g(x,\zeta) \quad \text{and} \quad
\widetilde{\cal Z}^{-1}[\tilde{g}](\zeta) = \int\limits_0^{2\pi}\frac{\d x}{2\pi}\tilde{g}(x,\zeta).
\end{equation*}
Defining the operator ${\cal U}[g](x,\zeta)\coloneqq \e{-\i x\zeta}g(x,\zeta)$, it is easy to show that $\widetilde{\cal Z}^{-1}{\cal U}{\cal Z}[f]=\sqrt{2\pi}{\cal F}[f]$ and hence related to the Fourier transformation. Likewise, one can also show that ${\cal Z}^{-1}{\cal U}^{-1}\widetilde{\cal Z}=\sqrt{2\pi}{\cal F}^{-1}(f)$. We just consider the case of the Fourier transform here, for which we have 
\begin{eqnarray*}
\widetilde{\cal Z}^{-1}{\cal U}{\cal Z}[f] &=&
\int\limits_0^{2\pi}\frac{\d x}{2\pi}\infsum{k}f(x+2\pi k)\es{-2\i\pi k\zeta}\es{-\i x\zeta} 
=\infsum{k}\int\limits_0^{2\pi}\frac{\d x}{2\pi}f(x+2\pi k)\es{-\i(x+ 2\pi k)\zeta} \\
&=&\frac{1}{2\pi}\int\limits_{\mathbbm{R}}\d x f(x)\es{-\i x\zeta} 
=
\sqrt{2\pi}{\cal F}[f](\zeta).
\end{eqnarray*}
Given this, we also have the following equality:
\begin{equation*}
{\cal U}{\cal Z}[f]=\sqrt{2\pi}\widetilde{Z}[{\cal F}[f]].    
\end{equation*}
Considering the explicit forms of ${\cal Z}$ and $\widetilde{\cal Z}$, we finally obtain
\be
\infsum{n}f(x+2\pi n)\es{-2\pi\i n\zeta}\es{-\i x\zeta} = \sqrt{2\pi}\infsum{n}{\cal F}[f](\zeta+n)\es{\i nx}.
\ee
If we now choose $\zeta=x=0$, we obtain the standard Poisson summation formula. As far as the application on complexifier coherent states is considered, we can choose $f\coloneqq g\circ S_t$, where $S_t$ is a scaling by the classicality parameter $t$ defined as $S_t x\coloneqq \frac{x}{\sqrt{t}}$, and we obtain the form of the Poisson summation formula used in this context. 
\section{Kummer functions as solutions of the heat equation}
\label{sec:SolHeat}
In this section, we briefly want to discuss the physical interpretation of Kummer functions in the context of self-similar solutions of the heat equation following closely the ideas of \cite{SelfSimHeat} and slightly generalising some aspects of their work. We start with the heat equation 
\begin{equation*}
\frac{\partial u}{\partial t}(x,t) = \kdiff\frac{\partial^2}{\partial x^2}u(x,t).
\end{equation*}
Along the lines of \cite{SelfSimHeat}, we introduce self-similar coordinates $(\xi,\tau)$ given by
\begin{equation*}
  \tau\coloneqq \tau(t) \quad \text{and} \quad   \xi\coloneqq \frac{x}{\sqrt{\kdiff}L(\tau)},
\end{equation*}
where the explicit form of the functions $L$ and $\tau$ still needs to be determined. As an ansatz for a self-smiliar solution of the heat equation, we consider
\begin{equation*}
u(x,t)=A(\tau(t))w(\xi(x,t),\tau(t))    ,
\end{equation*}
which leads to the following differential equation:
\begin{equation}
\label{eq:DeqnXiTau}
\dot{\tau} \left(\left(\frac{A'}{A}\right)w+\frac{\partial w}{\partial \tau }-\left(\frac{L'}{L}\right)\frac{\partial w}{\partial \xi}\right) = \frac{\kdiff}{\kdiff L^2(\tau)}\frac{\partial^2}{\partial\xi^2}w,   
\end{equation}
where the dot denotes a derivative with respect to $t$ and a prime one with respect to $\tau$. Now, the requirements made in \cite{SelfSimHeat} are that the self-similar solution is static in the $(\xi,\tau)$-frame, and hence there cannot be any explicit time-dependence. This yields a relation between $\tau$ and $L(\tau)$ of the form $\dot{\tau}\stackrel{!}{=}\frac{1}{L^2(\tau)}$. Furthermore, we are only interested in solutions for which $\frac{L'}{L}\eqqcolon G={\rm const}>0$ and $\frac{A'}{A}\eqqcolon \beta={\rm const}$.\footnote{Note that our $\beta$ is equal to $b$ in the notation of \cite{SelfSimHeat}.} As can be seen directly from (\ref{eq:DeqnXiTau}), the first condition corresponds to a constant expansion rate and the second condition determines the scaling of the amplitude $A(\tau(t))$ with $t$ parametrised by $b$. As shown in \cite{SelfSimHeat}, these conditions yield the following expressions for the functions $L$ and $A$ :
\begin{eqnarray}
L(\tau)& =
& L_0\e{G\tau}\quad \Leftrightarrow \quad L(t)=\sqrt{2G}\left(t-t_0\right)^{\frac{1}{2}} \quad \text{and} \nonumber\\
A(\tau) &=& A_0\e{\beta\tau}\quad \Leftrightarrow \quad A(t)=A_0\left(\frac{2G}{L_0^2}(t-t_0)\right)^{\frac{\beta}{2G}} .
\end{eqnarray}
We realise that the scaling of $L(t)$ is always with $t^{1/2}$, whereas the one of $A(t)$ can be different from $t^{-1/2}$, which holds for the heat kernel solution, for appropriate choices of $\frac{\beta}{G}$ --- as emphasised by the authors in \cite{SelfSimHeat}. With the above assumptions at hand, the differential equation that $w$ has to satisfy for a static self-similar solution of the heat equation in the $(\xi,\tau)$-frame reads
\begin{equation}
\label{eq:DeqnXiTau2}
\frac{\partial^2}{\partial \xi^2}w+G\xi\frac{\partial}{\partial\xi}w - \beta w =0    .
\end{equation}
Introducing the following scaling of the $\xi$-coordinate as well as the variable $W$
\begin{equation}
\tilde{\xi}\coloneqq \sqrt{\frac{G}{2}}\xi,\quad\quad W\coloneqq \e{\frac{\tilde{\xi^2}}{2}}w  ,  
\end{equation}
the differential equation for $w$ in (\ref{eq:DeqnXiTau2}) can be transformed into a generalised Hermite differential equation for $W$ given by
\begin{equation}
\label{eq:GenHerm}
\frac{\partial^2}{\partial\tilde{\xi}^2}W-2\tilde{\xi}\frac{\partial}{\partial\tilde{\xi}}W+2\tilde{\nu}W=0,\quad \tilde{\nu}=-\left(\frac{\beta}{G}+1\right).    
\end{equation}
In contrast to the standard Hermite differential equation, $\tilde{\nu}$ does not necessarily have to be a natural number. If we go a little beyond the discussion in \cite{SelfSimHeat} and perform a further final substitution of the variables according to $z\coloneqq \tilde{\xi}^2$ with $F(z)=F(\tilde{\xi}^2)$, then we can easily show that the differential equation in (\ref{eq:GenHerm}) transforms into Kummer's differential equation with the special choice of $b=\frac{1}{2}$ and $a=-\frac{\tilde{\nu}}{2}=\frac{\beta}{G}+1$:
\begin{equation}
z\frac{\d^2F}{\d z^2}+\left(\frac{1}{2}-z\right)\frac{\d F}{\d z}+\frac{\tilde{\nu}}{2}F=0    .
\end{equation}
The two linearly independent solutions are given by $\kchf{-\frac{\tilde{\nu}}{2},\frac{1}{2},z}$ and $U\left(-\frac{\tilde{\nu}}{2},\frac{1}{2},z\right)$. This allows to express the self-similar solution of the heat equation $u(x,t)$ in terms of Kummer functions of the first and second kind as
\begin{equation*}
u(x,t) = A(t;\tilde{\nu};G)
\es{-\frac{\xi^2G}{2}}\left(c_1(\tilde{\nu})U\left(-\frac{\tilde{\nu}}{2},\frac{1}{2},\frac{\xi^2G}{2}\right)
+ c_2(\tilde{\nu})\kchf{-\frac{\tilde{\nu}}{2},\frac{1}{2},\frac{\xi^2G}{2}}\right) ,
\end{equation*}
with
\begin{equation}
\label{eq:ANuG}
\tilde{\nu}\coloneqq -\left(\frac{\beta}{G}+1\right) , \quad A(t;\tilde{\nu};G)\coloneqq A_0\left(\frac{2G}{L_0^2}(t-t_0)\right)^{-\frac{1}{2}\left(\frac{\tilde{\nu}}{2}+1\right)} \quad \text{and} \quad \xi=\frac{x}{\sqrt{2\kdiff(t-t_0)}}  .
\end{equation}
In \cite{SelfSimHeat}, they do not perform the last transformation into the Kummer differential equation and thus this is probably the case why they do not relate the first independent solution to the Kummer function of the first kind, which automatically occurs in our discussion here. As discussed in \cite{SelfSimHeat}, in the special case where $\tilde{\nu}$ is an even integer the two Kummer functions are multiples of each other and can be identified with the Hermite polynomials and are no longer independent functions. In this case, next to the solution $U\left(-\frac{\tilde{\nu}}{2},\frac{1}{2},\frac{{\xi}^2}{2}\right)$ we can use  $\xi\kchf{\frac{1}{2}-\frac{\tilde{\nu}}{2},\frac{3}{2},\frac{{\xi}^2}{2}}$ as a second independent solution. 
\\
Interestingly, as far as $\e{-\frac{\xi^2G}{2}}\kchf{-\frac{\tilde{\nu}}{2},\frac{1}{2},\frac{\xi^2G}{2}}$ is considered, this is exactly the expression that we obtain in the computation of the semiclassical expectation values in subsection \ref{sec:QMdirect} and subsection \ref{sec:qU1}, respectively . Hence, the result of the Fourier transform involved in these computations corresponds to a self-similar solution of the $1+1$-dimensional heat equation. The fractional power $r$ of the momentum operator in these semiclassical expectation values determines the scaling behaviour of the time dependent amplitude of the self-similar solution. As can be easily seen and has been already discussed in \cite{SelfSimHeat}, for $\tilde{\nu}\not=0$ we obtain a scaling behaviour of the amplitude that differs from the standard $t^{-\frac{1}{2}}$. Carried over to the expectation values of fractional powers of the momentum operator, the case $\tilde{\nu}=0$ corresponds to the scenario where the operator becomes the identity operator and the expectation value the squared norm of the coherent states. For complexifier coherent states based on the analytic continuation of the heat kernel, this is the expected scaling behaviour of the norm with respect to the classicality parameter that can be identified with the temporal coordinate of the heat equation in this context. 
\section{Conclusions and Outlook}
\label{sec:Concl}
In this article, we extended former results of \cite{Tolar,Olmo,Kowalski_1996,Kastrup:2005xb,Bahr:2006kb} on coherent states on the circle in two different directions. We showed that we can compute semiclassical expectation values of fractional powers of momentum operators by means of Kummer functions and we have demonstrated this in section \ref{sec:QMCase} and \ref{sec:ZakTrafo} for $L_2(\mathbbm{R})$ and $L_2(S_1)$, respectively. For all operators considered in this work, the involved integrals could be computed exactly analytically without the need to perform any estimates within the calculations, as it has been done in \cite{GCS3} for fractional powers. Furthermore, since the asymptotic behaviour for Kummer functions is well-known in this context, it can be used to compute the expansion of these semiclassical expectation values in terms of the semiclassical parameter. It turns out that we automatically end up with the correct fractional power in the classical limit due to the fact that we do not need to estimate the integrals.

As a further result, we also discussed the computation of generic semiclassical matrix elements in the context of the Zak transformation and we were able to show that there exists a simple relation between semiclassical expectation values in $L_2(\mathbbm{R})$ and $L_2(S_1)$ --- as discussed in section \ref{sec:RelQMU1}. Given an operator $\hat{O}_{\rm QM}$  that is well-defined on the set of coherent states, we can compute the following associated matrix element  $\langle T_{2\pi n}\Psi^\hbar_{q',p'}\,|\, \hat{O}_{\rm QM}\,|\, \Pqp\rangle$ in $L_2(\mathbbm{R})$, where $T_{2\pi n}$ denotes a translation operator that translates by $2\pi n$ with $n\in\mathbbm{N}$. The matrix element for $L_2(S_1)$ can be expanded into a Fourier series whose Fourier coefficients $c_n$ are then exactly given by $c_n=\langle T_{2\pi n}\Psi^\hbar_{q',p'}\,|\, \hat{O}_{\rm QM}\,|\, \Pqp\rangle$. This shows that the semiclassical matrix elements in $L_2(S_1)$ are completely determined by the corresponding 'translated' matrix elements in $L_2(\mathbbm{R})$. The variable in which the Fourier transform is evaluated is exactly the parameter $\delta$ used in \cite{Bahr:2006kb,Kastrup:2005xb} that naturally enters the definition of the coherent states because it is the second argument of the Zak transform and for coherent states on the circle it can be understood as an additional fixed parameter in the interval 0 to 1. For a more detailed discussion on the physical properties of this parameter, see for instance \cite{Kastrup:2005xb}. Given this relation, semiclassical matrix elements like for instance in \cite{Olmo, Bahr:2006kb,Kowalski_1996,GCS3,Bahr:2006kb} can be computed in an alternative and possibly simpler manner. If we have an operator on $L_2(S_1)$ of which we want to compute semiclassical matrix elements, we just compute its Fourier coefficients, which in turn are simply matrix elements with respect to standard harmonic oscillator coherent states. Given these matrix elements, we can without any further computation directly write down the corresponding result for the matrix element in $L_2(S_1)$. 
In this way, we can avoid to perform explicitly the Poisson summation formula because this is automatically build into the formalism of the Zak transformation. This might likely reduce the actual effort of these semiclassical computations. Compared to the results of the expectation values in terms of Jacobi's theta functions and its derivatives as done in \cite{Olmo,Kowalski_1996, Kastrup:2005xb}, we believe that the relation from lemma \ref{lem:RelQMU1} in equation (\ref{eq:FinRelQMU1}) provides a more convenient alternative as far as the extraction of the classical limit is concerned. 

Having restricted our considerations to coherent states on the circle, a natural question is whether the techniques introduced here can be generalised to more complicated situations. As already mentioned before, if we consider the Zak transform as a map from $L_2(\mathbbm{R}^n)$ to $L_2(\mathbbm{R}^{2n}/\mathbbm{Z}^{2n})$, the relation of the matrix elements discussed in section \ref{sec:RelQMU1} carries over to the higher but finite dimensional case. As far as operators with fractional powers are concerned in a higher dimensional model, the operators can become more complicated functions of fractional powers than we considered here and thus it can happen that the integrals involved can no longer be solved by just using Kummer functions. However, as we discuss in a companion paper \cite{Giesel:2021yop}, similar techniques can be used for U$(1)^3$ coherent states and a certain class of dynamical operators (generlisations of the operator considered in \ref{sec:qU1}), also considered in \cite{Brunnemann1,Brunnemann2}, which improve the final semiclassical expansion in certain aspects. In the context of loop quantum gravity, a generalisation from U$(1)^3$ to SU(2) of this procedure would be comfortable to have at hand, in particular also because the semiclassical computations for SU(2) coherent states are much more involved in this case, so any simplification in this direction would be welcome. Since the theta function can also be defined for SU(2) \cite{ThetaSU2}, it is a starting point for analysing in more detail whether a Zak transformation or a generalisation thereof can be used for SU(2) coherent states in a similar way. We plan to investigate this in our future work. 
\section*{Acknowledgements}
K.G. thanks Jos\'{e} Mour\~{a}o for fruitful discussions on theta functions during the Jurekfest in Warsaw. 
The work of D.W. was supported by a stipend provided from the FAU Erlangen-Nürnberg. D.W. thanks the German Academic Scholarship Foundation for financial support in an earlier stage of this work.
\bibliography{CoherentStates}
\bibliographystyle{unsrt}
\end{document}